\documentclass[fleqn,usenatbib,usedcolumn]{mnras}
\let\la=\lesssim
\usepackage{graphicx}
\usepackage{amsmath}
\usepackage{amssymb}
\usepackage{bm}
\usepackage{tikz}
\usepackage{lipsum}
\usepackage{ulem}
\usetikzlibrary{shapes, shapes.geometric, arrows, arrows.meta, calc}

\newcommand{\fR}{\bar{f}_\mathrm{R0}}
\newcommand{\afR}{|\bar{f}_\mathrm{R0}|}
\newcommand{\vvir}{v_\mathrm{vir}}
\newcommand{\cvir}{c_\mathrm{vir}}
\newcommand{\Udisc}{\Upsilon_\mathrm{disc}}
\newcommand{\Ubulge}{\Upsilon_\mathrm{bulge}}
\newcommand{\vdisc}{v_\mathrm{disc}}

\definecolor{mygreen1}{RGB}{237, 248, 233}
\definecolor{mygreen2}{RGB}{186, 228, 179}
\definecolor{mygreen3}{RGB}{116, 196, 118}
\definecolor{mygreen4}{RGB}{49, 163, 84}
\definecolor{mygreen5}{RGB}{0, 109, 44}

\title[Rotation Curves in $f(R)$ Gravity]{Constraints on Chameleon $f(R)$-Gravity from Galaxy Rotation Curves of the SPARC Sample}

\author[A. P. Naik et al.]{Aneesh P. Naik,$^{1}$\thanks{E-mail: an485@ast.cam.ac.uk}
Ewald Puchwein,$^{1,2,3}$
Anne-Christine Davis,$^{4}$
\newauthor
Debora Sijacki,$^{1,2}$
and Harry Desmond$^{5}$
\\
$^{1}$Institute of Astronomy, University of Cambridge, Madingley Road, Cambridge, CB3 0HA, UK\\
$^{2}$Kavli Institute for Cosmology, University of Cambridge, Madingley Road, Cambridge, CB3 0HA, UK\\
$^{3}$Leibniz-Institut f{\"u}r Astrophysik Potsdam (AIP), An der Sternwarte 16, D-14482 Potsdam, Germany\\
$^{4}$Department of Applied Mathematics and Theoretical Physics, Centre for Mathematical Sciences, Cambridge CB2 0WA, UK\\
$^{5}$Astrophysics, University of Oxford, Denys Wilkinson Building, Keble Road, Oxford OX1 3RH, UK
}

\date{Accepted XXX. Received YYY; in original form ZZZ}
\pubyear{2019}

\begin{document}
\label{firstpage}
\pagerange{\pageref{firstpage}--\pageref{lastpage}}
\maketitle

% ABSTRACT
\begin{abstract}

In chameleon $f(R)$-gravity, the fifth force will lead to `upturns' in galaxy rotation curves near the screening radius. The location of the upturn depends on the cosmic background value of the scalar field $\fR$, as well as the mass, size and environment of the galaxy. We search for this signature of modified gravity in the SPARC sample of measured rotation curves, using an MCMC technique to derive constraints on $\fR$. Assuming NFW dark matter haloes and with $\fR$ freely varying for each galaxy, most galaxies prefer $f(R)$ gravity to $\Lambda$CDM, but there is a large spread of inferred $\fR$ values, inconsistent with a single global value. Requiring instead a consistent $\fR$ value for the whole sample, models with $\log_{10}\afR > -6.1$ are excluded. On the other hand, models in the range $-7.5<\log_{10}\afR<-6.5$ seem to be favoured with respect to $\Lambda$CDM, with a significant peak at -7. However, this signal is largely a result of galaxies for which the $f(R)$ signal is degenerate with the core/cusp problem, and when the NFW profile is replaced with a cored halo profile, $\Lambda$CDM gives better fits than any given $f(R)$ model. Thus, we find no convincing evidence of $f(R)$ gravity down to the level of $\afR \sim 6 \times 10^{-8}$, with the caveat that if cored halo density profiles cannot ultimately be explained within $\Lambda$CDM, a screened modified gravity theory could possibly provide an alternative solution for the core/cusp problem. However, the $f(R)$ models studied here fall short of achieving this.

\end{abstract}

% KEYWORDS

% INTRODUCTION
\section{Introduction}
\label{S:Intro}

In the previous two decades, the $\Lambda$CDM cosmological model has proven remarkably successful at describing the large-scale structure of the Universe \citep{Planck2016}. Despite this success, there remain some outstanding theoretical questions, such as the precise nature of dark matter, and the large discrepancy between (naively) expected and observed values of the cosmological constant $\Lambda$ \citep{Bull2016}.

In addition to these theoretical issues, there also remain a few unresolved tensions with observations on galactic scales, a review of which can be found in the article by \citet{Bullock2017}. There are three problems in particular that have been the subjects of widespread attention. Firstly, the `missing satellites' problem \citep{Moore1999, Klypin1999}, which arises from a mismatch between the observed number of satellite galaxies of the Milky Way and that expected from the number of substructures in $\Lambda$CDM cosmological simulations. Secondly, the `too-big-to-fail' problem, which arises from the unexpectedly low central densities of observed Milky Way satellites \citep{BoylanKolchin2011}.

The (likely related) third problem, the so-called `core/cusp' problem, is of particular relevance here. A key prediction of dark matter-only $\Lambda$CDM simulations is that of `cuspy' density profiles for dark matter haloes, i.e. profiles with steep central slopes \citep{Navarro1997}, roughly $\propto r^{-1}$. On the other hand, some observations \citep{Moore1994, Flores1994, Walker2011, Oh2015}, e.g., of galaxy rotation curves, have been found to suggest flatter, `cored' profiles in the inner regions of haloes.
As a possible resolution, some simulations including the effects of baryonic feedback have suggested that such mechanisms can induce cored central density distributions in dark matter haloes \citep{Mashchenko2008, Governato2010, Pontzen2012, DiCintio2014}. This is thought to happen when feedback, e.g., from supernovae pushes around large amounts of gas in the central region of a galaxy, the resulting fluctuation in the gravitational potential of the gas could then also affect the dark matter distribution. Other simulations, however, find no such effect \citep[e.g,][]{Bose2018}. In addition, the recent work of \citet{Genina2018} found that the inference of a cored profile from observations might be the result of an incorrect assumption of spherical symmetry, and that observed galaxies are consistent with cusps. The core/cusp debate hence remains far from resolved.

Because of these theoretical and observational tensions, alternatives and extensions to $\Lambda$CDM warrant investigation. There have been many works investigating viable substitutes for cold dark matter, such as `self-interacting dark matter' which has been shown to produce cored haloes in simulations \citep{Vogelsberger2012}. Alternatively, a generic and widely-adopted method of extending $\Lambda$CDM is through the introduction of a scalar field coupled to gravity, giving rise to gravitational-strength couplings to matter: so-called `fifth forces'. Overviews of these modified gravity theories can be found in, e.g. the works by \citet{Amendola2010, Clifton2012, Joyce2015, Koyama2016}.

For the scalar fields to retain cosmological relevance, they need sufficiently low masses (i.e. large Compton wavelengths), but their fifth forces will consequently violate stringent Solar System tests of gravity unless a `screening mechanism' is introduced \citep{Jain2010, Khoury2010}. Most relevant to the present work is the chameleon mechanism \citep{Khoury2004}, in which the mass of the scalaron increases in dense environments, where the fifth force is consequently rendered undetectable.

The theory investigated in this work  falls within the widely-studied class of $f(R)$ gravity models. First described in \citet{Buchdahl1970}, the idea behind $f(R)$ gravity is simple: the Ricci scalar $R$ in the Einstein-Hilbert action is generalised to $R+f(R)$. The Hu-Sawicki form of $f(R)$ \citep{Hu2007} has been shown to give a theory that is formally equivalent to a subclass of chameleon theories \citep{Brax2008}. The theory has one key parameter, the cosmic background value of the scalar field, $\fR$, constraints on which can therefore be translated into constraints on generalised chameleon parameters.

It is worth noting that the significant recent `multi-messenger' observation of GW170817 placed severe constraints on many modified gravity theories, which posited differing speeds for gravitational and electromagnetic waves. $f(R)$ gravity, however, is consistent with such constraints, predicting a gravitational wave speed equal to that of light \citep{Lombriser2017, Sakstein2017, Maria2017}.

Chameleon gravity has been tested using a multitude of methods, on a range of length scales from the laboratory to the CMB. Review articles summarising these tests and corresponding constraints can be found in the works by \citet{Lombriser2014, Burrage2016, Burrage2018}. In particular, astrophysical tests have been found to give some of the strongest constraints on $f(R)$ gravity to date. For instance, distances inferred using Cepheid variables and stars at the tip of the Red Giant Branch would be different under modified gravity than those in $\Lambda$CDM, and can therefore be used to constrain modified gravity theories. Using such a method, \citet{Jain2013} find an upper bound of $\afR=5\times 10^{-7}$. As another example, constraints at around the $10^{-6}$ level were found by \citet{Xu2015}, considering the effect of $f(R)$ gravity on redshift-space distortions. A similar upper bound was calculated by \citet{Vikram2018}, who searched for evidence of significant differences in the gaseous and stellar rotation curves of isolated dwarf galaxies; the self-screening of stars would mean that in contrast to the gas, there would be no enhancement of the stellar rotation velocity due to the fifth force. 

\citet{Vikram2013} look for signatures of displacement between galactic stellar and gaseous components, and signatures of warping as a result of this displacement, but find no significant constraints on modified gravity. Employing similar principles, \citet{Desmond2019, Desmond2018b} use updated methods and data to search for these stellar-gas offsets, calculating an upper bound on $\afR$ of a few $\times 10^{-8}$. \citet{Desmond2018a}, meanwhile, searches for the warping signature. Interestingly, both of these studies independently find a possible fifth force signal. In the language of these studies, which vary the fractional change in effective Newton's constant $\Delta G/G_N$ and Compton wavelength $\lambda_C$, there is a signal at $\Delta G/G_N \approx 10^{-2}$ and $\lambda_C \approx 2$ Mpc. This is irrelevant to $f(R)$, which predicts $\Delta G/G_N = 1/3$ in unscreened regions, but interesting nonetheless. However, the authors add the cautionary note that the signal could well be a result of a number of other effects, including unaccounted-for galaxy formation physics.   

In \citet{Naik2018}---henceforth Paper I---it was found that the presence of a chameleon fifth force would lead to observable imprints on galactic rotation curves. Specifically, there will be a `screening radius' beyond which the fifth force is active. Within the radius, standard gravity is restored. This has the consequence that an `upturn' appears in the rotation curve at this screening radius, with a fifth force enchancement beyond the screening radius. The location of this screening radius within a given galaxy depends on the global value of $\fR$ and the environment of the galaxy, as well as its mass and size.

In this work, we use the SPARC sample of high-quality rotation curve measurements \citep{Lelli2016} to search for this signature. After several cuts to the sample, 85 galaxies remain, each of which we model with a gaseous disc, a stellar disc, a dark matter halo, and where appropriate, a stellar bulge. Using these components, we solve the $f(R)$ scalar field equations for the fifth force contribution, thereby constructing a forward model for the galaxy's rotation curve. With these models, we then employ an MCMC technique to explore the posterior probabilities for the parameters for each galaxy, looking in particular at the inferred values for the background value of the scalar field $\fR$. We also perform model comparisons between $f(R)$ and $\Lambda$CDM using the likelihood ratios calculated for our best fit models.

Several recent studies have worked with the SPARC sample, e.g. \citet{McGaugh2016, Desmond2017a, Desmond2017b, Lelli2017, Li2018}. Of particular relevance to the present work, however, is the study of \citet{Katz2017}. In that work, the SPARC galaxies were used to investigate the core/cusp problem. It was found in that work that fits to the rotation curves typically improved when cuspy `NFW' haloes \citep{Navarro1997} were replaced with the cored `DC14' haloes of \citet{DiCintio2014}, derived empirically from $\Lambda$CDM simulations incorporating subgrid stellar feedback mechanisms.

We argued in Paper I that the NFW profile provides not only a good fit to dark matter haloes in $\Lambda$CDM simulations, but also in $f(R)$ simulations. In fact in $f(R)$ gravity, depending on halo mass, a higher concentration of the NFW profile is needed \citep[e.g.][]{Arnold2019}. Thus, $f(R)$ gravity does not reduce the central densities of haloes and consequently does not offer a direct solution to the core/cusp problem. However, \citet{Lombriser2015} made an interesting observation: a chameleon fifth force in the outer regions of a cuspy halo can mimic the observational signature of a cored halo profile in measurements of the kinematics of galaxies, we will discuss this possibility further in \S\ref{S:Results} and \S\ref{S:Conclusions}.

Baryonic feedback mechanisms that could in principle lead to the formation of a halo core, would likely be as able to do so under modified gravity as under $\Lambda$CDM. We, therefore explore the impact of a cored density profile by computing results both for NFW and DC14 haloes \citep[similar to][]{Katz2017}. We utilise a model in which $\fR$ is allowed to vary freely for each galaxy, as well as a grid of models covering a range of $\fR$ values fixed for the whole sample. In the former case, we analyse the marginal posteriors on $\fR$ for each galaxy to find any evidence of significant clustering around a single preferred value. In the latter case, we are able to compare each model with $\Lambda$CDM, and thus search more directly for any evidence of a globally preferred value for $\fR$.

We also explore several different prescriptions for the mass-to-light ratios, and the impacts of stellar and environmental screening.

All of the code, analysis tools, and plotting scripts used in this work have been made into a publicly available \texttt{python 3} package\footnote{\href{https://github.com/aneeshnaik/spam}{https://github.com/aneeshnaik/spam}}.

This work is structured as follows. Firstly, \S\ref{S:Theory} briefly describes $f(R)$ gravity, providing the theoretical details relevant to this work. Next, \S\ref{S:Data} describes the SPARC data set and \S\ref{S:Methods} describes the methodology of the present work. Results are then outlined in \S\ref{S:Results}, before discussion and concluding remarks in \S\ref{S:Conclusions}.

% THEORY
\section{$\lowercase{f}(R)$ Gravity}
\label{S:Theory}

In this section, we summarise the most salient theoretical details about $f(R)$ gravity. A more in-depth description of the relevant aspects can be found in Paper I and the references therein.

$f(R)$ gravity was first introduced in \citet{Buchdahl1970}, more recent reviews can be found in the works of \citet{Amendola2010, Clifton2012, Joyce2015, Koyama2016}.

%As mentioned in the Introduction, the original $f(R)$ work can be found in the article by \citet{Buchdahl1970}, and excellent reviews can be found in the works of \citet{Amendola2010, Clifton2012, Joyce2015, Koyama2016}.

The action of this theory is given by:
\begin{equation}\label{E:f(R)Action}
    S = \int d^4x\sqrt{-g}\frac{1}{16\pi G}\left[R+f\left(R\right)\right] + S_{m}[g_{\mu\nu},\psi_{i}],
\end{equation}
where $S_m$ is the matter action, governing the matter fields $\psi_i$. As the name of the theory would suggest, the Ricci scalar $R$ in the classical Einstein-Hilbert action is replaced with a generalised $R+f\left(R\right)$. The theory reduces to $\Lambda$CDM in the case that $f(R)=-2\Lambda$.

In $f(R)$ gravity, the derivative $f_\mathrm{R} \equiv \frac{\mathrm{d}f}{\mathrm{dR}}$ plays the role of a scalar degree of freedom of the theory, similar to a scalar field. $f_\mathrm{R}$ is governed by the equation
\begin{equation}\label{E:FieldEOM}
    \nabla^2 f_\mathrm{R} = \frac{1}{3c^2} \left( \delta R - 8\pi G \delta \rho \right),
\end{equation}
which can be derived by extremising the action in Eq.~(\ref{E:f(R)Action}) with respect to the metric, and making a quasistatic approximation \citep{Sawicki2015, Noller2014}. Here $\delta\rho$ is the perturbation of the matter density from its background value, while $\delta R$ denotes the perturbation of the scalar curvature.

The specific $f(R)$ model we consider in this work is that of \citet{Hu2007}, which is known to exhibit a `chameleon' screening mechanism, whereby gradients of the scalar field are suppressed in dense regions. Equivalently, this can be framed as the mass of the scalaron being environment dependent, and growing unobservably large in dense environments. In these `screened' regions, the range of the fifth force plummets, rendering it unobservable, and thus allowing such theories to escape stringent laboratory and Solar System constraints. In this theory, $\delta R$ is given by
\begin{equation}\label{E:CurvaturePerturbation}
\delta R = \bar{R}(a)\left(\sqrt{\frac{\bar{f}_\mathrm{R}(a)}{f_\mathrm{R}}}-1\right),
\end{equation}
where $\bar{f}_\mathrm{R}(a)$ and $\bar{R}(a)$ are the background values of the scalar field and scalar curvature respectively, at a given cosmic scale factor $a$. These are given in turn by
\begin{equation}\label{E:BackgroundFieldTimeDependence}
\bar{f}_\mathrm{R}(a) = a^6\fR\left(\frac{1 + 4\frac{\Omega_\Lambda}{\Omega_m}}{1 + 4\frac{\Omega_\Lambda a^3}{\Omega_m}}\right)^2,
\end{equation}
\begin{equation}\label{E:BackgroundCurvature}
\bar{R}(a) = 3\frac{H_0^2\Omega_m}{a^3}\left(1 + 4\frac{\Omega_\Lambda a^3}{\Omega_m}\right).
\end{equation}

Equation (\ref{E:FieldEOM}), combined with (\ref{E:CurvaturePerturbation}), (\ref{E:BackgroundFieldTimeDependence}), and (\ref{E:BackgroundCurvature}), can be used to solve for the scalar field $f_R$ everywhere. The only inputs needed are the mass distribution (via the density perturbation $\delta\rho$) and the cosmic background value of the scalar field $\fR$. 

The fifth force, i.e. the resultant deviation from standard gravity, is then given by gradients of the scalar field $f_R$. The corresponding acceleration reads
\begin{equation}
\label{E:a5}
    \bm{a_5} = \frac{1}{2}c^2\bm{\nabla} f_\mathrm{R}.
\end{equation}

In Paper I, it was found by post-processing simulations that for certain values of $\fR$, galaxies exhibit screening radii, beyond which rotation curves would show enhancements due to this fifth force.

% DATA
\section{Data}
\label{S:Data}

    For this investigation, we have used the rotation curves from the SPARC sample \citep{Lelli2016}. The SPARC sample consists of 175 high-quality HI/H$\alpha$ rotation curves, along with accompanying $3.6\mu m$ photometry from Spitzer. The sample is diverse, spanning 5 orders of magnitude in mass, and encompassing a variety of morphologies. $3.6\mu m$ surface brightness is believed to be a good tracer of stellar mass, so its pairing with rotation curve data enables detailed modelling of the different components of a given galaxy.

    For each galaxy, the SPARC sample includes a wealth of information. Most relevant is $v_\mathrm{circ}(R)$, the rotation speed as a function of radius, along with its corresponding error bars. In addition to this, the photometry information is converted into contributions to the rotation speed due to the stellar disc and stellar bulge, $v_\mathrm{disc}$ and $v_\mathrm{bulge}$. The bulge component is only included for galaxies for which the photometry profile departs significantly from an exponential disc profile in the central regions, a minority of the overall sample (13 of our final 85 galaxies). The quantities $v_\mathrm{disc}$ and $v_\mathrm{bulge}$ are based on a mass-to-light ratio of $1 \, M_\odot/L_\odot$, and therefore need re-scaling for any other assumed ratio. Finally, for each galaxy, gas surface brightness information has been converted into a gas contribution $v_\mathrm{gas}$. A more detailed description of the data and the derivation of the velocity components can be found in \citet{Lelli2016}.

    From the original sample of 175 galaxies, we make a series of cuts, yielding a final sample of 85 galaxies. The cuts are summarised below, along with numbers in parenthesis indicating how many galaxies were removed at each successive cut. 

    We include galaxies for which:
    \begin{itemize}
        \item There are $\geq$ 5 data points. $(175-4=171)$
        \item Inclination $i\geq 30^\circ.$ $(171-12=159)$
        \item Quality flag $Q = 1,2$. This excludes galaxies with major asymmetries, for which $Q=3$. $(159-10=149)$
        \item Information about the HI gas distribution is available. This excludes galaxies for which the $v_\mathrm{gas}$ contribution can only be estimated approximately. $(149-2=147)$
        \item There is not a significant environmental screening effect. See \S\ref{S:Methods:EnvScreening} for a more detailed description of this cut. $(147-62=85)$
    \end{itemize}

% METHODS
\begin{table*}
    \caption{Summary of the models investigated in this work. Columns are (1) Model: the alphabetical/alphanumerical ID given in this work to a given model, (2) Theory: $\Lambda$CDM or $f(R)$, i.e. whether or not a fifth force contribution is included in the rotation curve model, (3) $\fR$: for an $f(R)$ model, $\fR$ is either left as a freely `varying' parameter, or in the case of Models H0-19 and I0-19, imposed as a series of 20 fixed values given in Table \ref{T:HVals} (see discussion in \S\ref{S:Methods:Models}), (4) DM Halo:, NFW or DC14 dark matter halo profile (\S\ref{S:Methods:Models}), (5) Large-scale Env.: whether a large-scale environment is added to the scalar field solver (\S\ref{S:Methods:EnvScreening}). Note that in all models, we take into account environmental screening effects with the final sample cut discussed in \S\ref{S:Data}, and this additional test is merely to ensure that the remaining sample is largely unaffected by environmental screening. (6) Stellar Screening: whether the stellar component is \textit{excluded} from sourcing a fifth force in the scalar field solver (\S\ref{S:Methods:StellarScreening}), (7) $\Upsilon$: Whether the mass-to-light ratio is treated with one free parameter (using the same ratio for disc and bulge),`1', two free parameters (allowing for different values for disc and bulge),`2', or as `fixed' empirical values,`0' (\S\ref{S:Methods:Models}), and (8) Parameters: a list of the free parameters for a given model (see main text for details).}
    \begin{tabular}{c | ccccccc}
    Model & Theory       & $\fR$             & DM Halo & Large-scale Env. & Stellar Screening & $\Upsilon$ & Parameters                                               \\ \hline
    A     & $\Lambda$CDM & -                 & NFW     & -                & -                 & 1          & $\vvir$, $\cvir$, $\Upsilon$, $\sigma_g$                 \\
    B     & $f(R)$       & Varying           & NFW     & $\times$         & $\times$          & 1          & $\fR$, $\vvir$, $\cvir$, $\Upsilon$, $\sigma_g$          \\
    C     & $f(R)$       & Varying           & NFW     & \checkmark       & $\times$          & 1          & $\fR$, $\vvir$, $\cvir$, $\Upsilon$, $\sigma_g$          \\
    D     & $f(R)$       & Varying           & NFW     & $\times$         & \checkmark        & 1          & $\fR$, $\vvir$, $\cvir$, $\Upsilon$, $\sigma_g$          \\
    E     & $f(R)$       & Varying           & NFW     & $\times$         & $\times$          & 0          & $\fR$, $\vvir$, $\cvir$, $\sigma_g$                      \\
    F     & $f(R)$       & Varying           & NFW     & $\times$         & $\times$          & 2          & $\fR$, $\vvir$, $\cvir$, $\Udisc$, $\Ubulge$, $\sigma_g$ \\
    G     & $\Lambda$CDM & -                 & DC14    & -                & -                 & 1          & $\vvir$, $\cvir$, $\Upsilon$, $\sigma_g$                 \\
    H0-19 & $f(R)$       & 20 imposed values & NFW     & $\times$         & $\times$          & 1          & $\vvir$, $\cvir$, $\Upsilon$, $\sigma_g$                 \\
    I0-19 & $f(R)$       & 20 imposed values & DC14    & $\times$         & $\times$          & 1          & $\vvir$, $\cvir$, $\Upsilon$, $\sigma_g$
    \end{tabular}
    \label{T:Models}
\end{table*}

\section{Methods}
\label{S:Methods}

\subsection{Rotation Curve Models}
\label{S:Methods:Models}

In this study, we have experimented with a range of different models for galaxy rotation curves, e.g., models with different dark matter halo profiles, fixed or varying $\fR$, as well as different treatments of mass-to-light ratio, environmental effects and a potential self-screening of stars. A summary of the models investigated can be found in Table \ref{T:Models}. The terms used in this table will be explained over the course of the following few subsections.

For each galaxy, we construct a circular velocity model $v_\mathrm{model}$, given by
\begin{equation}\label{E:v_model}
    v_{\mathrm{model}}^2(r) = a_\mathrm{tot}r = (a_\mathrm{N}+a_5)r,
\end{equation}
where $r$ is the distance in the disc plane from the centre of the galaxy, $a_\mathrm{N}$ is the standard gravity (Newtonian) acceleration, and $a_5$ is the additional acceleration due to the fifth force.

The Newtonian acceleration $a_\mathrm{N}$ is sourced by the various components of the galaxy: the gas, stellar disc, stellar bulge, and dark matter. This relationship can be written as
\begin{equation}\label{E:a_Newton}
    a_\mathrm{N}(r) = \frac{ v_{\mathrm{gas}}^2 + v_{\mathrm{DM}}^2 + \Upsilon_\mathrm{disc}v_{\mathrm{disc}}^2 + \Upsilon_\mathrm{bulge}v_{\mathrm{bulge}}^2}{r},
\end{equation}
where $v_x$ is the contribution of component $x$ to the overall velocity curve as a function of radius $r$, and $\Upsilon_\mathrm{disc}$ and $\Upsilon_\mathrm{bulge}$ are the mass-to-light ratios for the stellar disc and bulge. The gaseous and stellar contributions ($v_\mathrm{gas}$, $v_\mathrm{disc}$, and $v_\mathrm{bulge}$) have been calculated by the SPARC team and are included within the SPARC data. Note also that the bulge term only applies to galaxies for which the SPARC team have found evidence for a stellar bulge, and have included a corresponding $v_\mathrm{bulge}$ contribution in the data; this corresponds to 13 of the 85 galaxies in our final sample.

All that remains to complete the model is to calculate the dark matter contribution $v_\mathrm{DM}$, the mass-to-light ratios $\Upsilon$, and the acceleration due to the fifth force $a_5$.

Firstly, we calculate the dark matter contribution $v_\mathrm{DM}$ by modelling the dark matter halo as a 2-parameter profile. We consider two different choices for the profile: the widely used Navarro-Frenk-White (NFW) profile \citep{Navarro1997}, as well as the `DC14' profile of \citet{DiCintio2014}. The latter profile is derived empirically from $\Lambda$CDM simulations incorporating baryonic feedback, and provides a more `cored' profile than NFW for galaxies with a significant stellar component. It was found by \citet{Katz2017} to provide a better fit to the SPARC galaxies than NFW. As indicated in Table \ref{T:Models}, Models A-F and H0-19 all use NFW haloes, while Models G and I0-19 employ DC14 haloes.

In either case, the profile is determined by 2 parameters, which we adopt as free parameters in our fit: the concentration parameter $c_\mathrm{vir}$ and the virial velocity $v_\mathrm{vir}$. $c_\mathrm{vir}$ is defined by
\begin{equation}
    c_\mathrm{vir} = \frac{r_\mathrm{vir}}{r_{-2}},
\end{equation}
where $r_\mathrm{vir}$ is the virial radius, the radius enclosing a region of average density 93.6 times the cosmic critical density, and $r_{-2}$ is the radius at which the logarithmic slope of the halo density profile is -2, which in the case of NFW is the same as the scale radius. The virial velocity, $v_\mathrm{vir}$ is in turn given by
\begin{equation}
    v_\mathrm{vir} = \sqrt{\frac{GM_\mathrm{vir}}{r_\mathrm{vir}}},
\end{equation}
where $M_\mathrm{vir}$ is the virial mass: the mass contained within the virial radius, approximated here by the total dark matter mass contained within the virial radius. Details of the translation between these parameters and $v_\mathrm{DM}$ can be found in, e.g., the appendix of \citet{Katz2018}.

Secondly, we have experimented with three approaches for the mass-to-light ratios $\Upsilon_\mathrm{disc}$ and $\Upsilon_\mathrm{bulge}$. In the most general case, we treat them as two free parameters for each galaxy. Alternatively, we treat them as a single free parameter (taking $\Upsilon_\mathrm{disc}=\Upsilon_\mathrm{bulge}$), mirroring the approach of \citet{Katz2017}. The final option is to treat them as fixed values ($\Udisc=0.5 M_\odot/L_\odot$ and $\Ubulge=0.7 M_\odot/L_\odot$). This was the approach taken in \citet{McGaugh2016}. As indicated in Table \ref{T:Models}, we mostly adopt the middle approach, that of a single free parameter (denoted as `1' in the $\Upsilon$ column of Table \ref{T:Models}). However, Model E takes the fixed values for the mass-to-light ratios (denoted as `0'), while Model F treats them as two free parameters (denoted as `2'). As mentioned previously, only 13 of the galaxies have a bulge component. For the remaining 72 galaxies, Models F and B are identical.

Finally, the fifth force contribution, $a_5$, is given by gradients of the $f(R)$ scalar field, $f_R$, via Eq. (\ref{E:a5}). In order to compute this, the scalar field $f_R$ needs to be calculated across the galactic disc by solving Equation (\ref{E:FieldEOM}). For this purpose, we use a spherical 1D scalar field solver, which takes as an input the mass distribution and the parameter $\fR$. The details of this solver are discussed in the next subsection.

\begin{table}
    \caption{Values of $\log_{10}\afR$ used in Models H0-19 and I0-19. They form a sequence of values that are uniformly spaced in the logarithm from $\log_{10}\left(1.563 \times 10^{-8}\right)$ to $\log_{10}\left(2 \times 10^{-6}\right)$.}
    \begin{tabular}{c | c || c | c}
    Model      & $\log_{10}|\fR|$ & Model & $\log_{10}|\fR|$ \\ \hline
    H0 (or I0) & -7.806         & H10   & -6.697         \\
    H1         & -7.695         & H11   & -6.586         \\
    H2         & -7.584         & H12   & -6.475         \\
    H3         & -7.473         & H13   & -6.364         \\
    H4         & -7.362         & H14   & -6.253         \\
    H5         & -7.252         & H15   & -6.143         \\
    H6         & -7.141         & H16   & -6.032         \\
    H7         & -7.030         & H17   & -5.921         \\
    H8         & -6.919         & H18   & -5.810         \\
    H9         & -6.808         & H19   & -5.699
    \end{tabular}
    \label{T:HVals}
\end{table}

As indicated in Table \ref{T:Models}, we adopt two different, but complementary approaches for the parameter $\fR$:

\begin{itemize}
    \item $\fR$ is a free parameter in the fit of the rotation curve for each galaxy. The resulting spread of marginal $\fR$ posteriors values across the galaxies of the sample can then be inspected for any significant clustering around a single preferred global value. This option is denoted as `varying' under the $\fR$ column in Table \ref{T:Models}, and is used for all $f(R)$ models except H0-19 and I0-19.
    \item By contrast, Models H0-19 and I0-19 take a log-space grid of 20 values of $\afR$, ranging from $\log_{10}\left(1.563 \times 10^{-8}\right)$ to $\log_{10}\left(2 \times 10^{-6}\right)$. These values are given in Table \ref{T:HVals}. Each of these 20 values are in turn globally imposed over the whole sample, for example Model H17 imposes a global $\log_{10}\afR$ value of -5.921. These 20 models can then be compared with each other and with $\Lambda$CDM, providing constraints on $\fR$.
\end{itemize}

\begin{figure*}
    % arrow style
\tikzstyle{arrow} = [thick,->,>=stealth]

% styles for main boxes
\tikzstyle{parameter} = [rectangle, rounded corners, minimum width=1.5cm, minimum height=1cm,text centered, draw=black, fill=mygreen1]
\tikzstyle{model} = [ellipse, minimum width=1.5cm, minimum height=1cm,text centered, draw=black, fill=mygreen2]
\tikzstyle{data} = [diamond, minimum width=1.5cm, minimum height=1cm,text centered, draw=black, fill=mygreen3]
\tikzstyle{calculated} = [rectangle, rounded corners, minimum width=1.5cm, minimum height=1cm,text centered, draw=black, fill=mygreen4]

% style for line intersections
\tikzstyle{junction} = [circle, fill=mygreen5, inner sep=0pt, minimum size=6pt]

% smaller versions of main box styles for legend handles
\tikzstyle{legend1} = [rectangle, rounded corners, minimum width=1.2cm, minimum height=0.8cm,text centered, draw=black, fill=mygreen1]
\tikzstyle{legend2} = [ellipse, minimum width=1.2cm, minimum height=0.8cm,text centered, draw=black, fill=mygreen2]
\tikzstyle{legend3} = [diamond, minimum width=1.2cm, minimum height=0.8cm,text centered, draw=black, fill=mygreen3]
\tikzstyle{legend4} = [rectangle, rounded corners, minimum width=1.2cm, minimum height=0.8cm,text centered, draw=black, fill=mygreen4]

\begin{tikzpicture}[node distance=2cm]

% posterior, top and centre of diagram
\node (posterior) [calculated, align=center] {Posterior $\mathcal{P}$ \\ Eq. (\ref{E:Posterior})};

% legend above posterior
\node (legend1) [legend1, above of=posterior, yshift=-0.5cm, align=center, label=Parameter] {};
\node (legend2) [legend2, right of=legend1, align=center, label=Tool or model] {};
\node (legend3) [legend3, right of=legend2, align=center, label=Data] {};
\node (legend4) [legend4, right of=legend3, align=center, label=Calculation] {};

% likelihood and prior boxes
\node (likelihood) [calculated, below of=posterior, xshift=-4cm, yshift=1cm, align=center] {Likelihood $\mathcal{L}$ \\ Eq. ($\ref{E:Likelihood}$)};
\node (prior) [calculated, below of=posterior, xshift=4cm, yshift=1cm, align=center] {Prior $\pi$ \\ \S\ref{S:Methods:Priors}};

% node connecting likelihood and prior to posterior
\node (node1) [junction, below of=posterior, yshift=1cm] {};
\draw [arrow] (likelihood) -- (node1) -- (posterior);
\draw [arrow] (prior)  -- (node1) -- (posterior);

% vdata and vmodel boxes
\node (vdata) [data, below of=likelihood, xshift=-2cm, yshift=0cm, align=center] {$v_\mathrm{data}$ \\ \S\ref{S:Data}};
\node (vmodel) [calculated, below of=likelihood, xshift=2cm, yshift=0cm, align=center] {$v_\mathrm{model}$ \\ Eq. ($\ref{E:v_model}$)};

% node connecting vdata, vmodel, sigma_g to likelihood
\node (node2) [junction, below of=likelihood, yshift=1cm] {};
\node (node3) [junction, below of=likelihood, yshift=1cm, xshift=2cm] {};
\draw [arrow] (vdata) |- (node2) -- (likelihood);
\draw [arrow] (vmodel) -- (node3) -- (node2) -- (likelihood);

% a5 and aN boxes
\node (a5) [calculated, below of=vmodel, xshift=-4cm, yshift=0.5cm, align=center] {$a_5$ \\ Eq. (\ref{E:a5})};
\node (aN) [calculated, below of=vmodel, xshift=4cm, yshift=0.5cm, align=center] {$a_N$ \\ Eq. (\ref{E:a_Newton})};

% node connecting a5 and aN to vmodel
\node (node4) [junction, below of=vmodel, yshift=0.5cm] {};
\draw [arrow] (aN) -- (node4) -- (vmodel);
\draw [arrow] (a5) -- (node4) -- (vmodel);

% fR solver box; connect to a5
\node (fRSolver) [model, below of=a5, yshift=0.5cm, align=center] {$f(R)$ solver \\ \S\ref{S:Methods:fRSolver}};
\draw [arrow] (fRSolver) -- (a5);

% rho boxes and nodes above them and arrows connecting it all
\node (rhoenv) [calculated, below of=fRSolver, xshift=2cm, yshift=1cm, align=center] {$\rho_\mathrm{env}$};
\node (node5) [junction, above of=rhoenv, yshift=-1cm] {};
\draw [arrow] (rhoenv) -- (node5) -- (fRSolver);
\node (rhoDM) [calculated, right of=rhoenv, align=center] {$\rho_\mathrm{DM}$};
\node (node6) [junction, above of=rhoDM, yshift=-1cm] {};
\draw [arrow] (rhoDM) -- (node6) -- (node5) -- (fRSolver);
\node (rhogas) [calculated, right of=rhoDM, align=center] {$\rho_\mathrm{gas}$};
\node (node7) [junction, above of=rhogas, yshift=-1cm] {};
\draw [arrow] (rhogas) -- (node7) -- (node6) -- (node5) -- (fRSolver);
\node (rhodisc) [calculated, right of=rhogas, align=center] {$\rho_\mathrm{disc}$};
\node (node8) [junction, above of=rhodisc, yshift=-1cm] {};
\draw [arrow] (rhodisc) -- (node8) -- (node7) -- (node6) -- (node5) -- (fRSolver);
\node (rhobulge) [calculated, right of=rhodisc, align=center] {$\rho_\mathrm{bulge}$};
\draw [arrow] (rhobulge) |- (node8) -- (node7) -- (node6) -- (node5) -- (fRSolver);

% phiext box; connect to rhoenv
\node (phiext) [calculated, below of=rhoenv, align=center] {$\Phi_\mathrm{ext}$};
\draw [arrow] (phiext) -- (rhoenv);

% screening map box; connect to phiext
\node (screeningmap) [model, below of=phiext, align=center] {screening map \\ \S\ref{S:Methods:EnvScreening}};
\draw [arrow] (screeningmap) -- (phiext);

% baryonic model boxes; connect to rho boxes
\node (gasdisc) [model, below of=rhogas, align=center] {exp. disc \\ \S\ref{S:Methods:fRSolver}};
\draw [arrow] (gasdisc) -- (rhogas);
\node (stellardisc) [model, below of=rhodisc, align=center] {exp. disc \\ \S\ref{S:Methods:fRSolver}};
\draw [arrow] (stellardisc) -- (rhodisc);
\node (stellarbulge) [model, below of=rhobulge, align=center] {Hernquist \\ \S\ref{S:Methods:fRSolver}};
\draw [arrow] (stellarbulge) -- (rhobulge);

% DM halo box; connect to rhoDM
\node (DMhalo) [model, below of=screeningmap, xshift=-0.5cm, align=center] {DM halo \\ \S\ref{S:Methods:Models}};
\coordinate (p3) at ($(screeningmap)+(2cm,0)$);
\draw [arrow] (DMhalo) -- (p3) -- (rhoDM);

% v component boxes
\node (vDM) [calculated, below of=stellardisc, xshift=-4cm, yshift=-2cm, align=center] {$v_\mathrm{DM}$ \\ \S\ref{S:Methods:Models}};
\node (vgas) [data, right of=vDM, align=center] {$v_\mathrm{gas}$};
\node (vdisc) [data, right of=vgas, xshift=0cm, align=center] {$v_\mathrm{disc}$};
\node (vbulge) [data, right of=vdisc, xshift=0cm, align=center] {$v_\mathrm{bulge}$};

% connect v boxes to aN
\node (node9) [junction, right of=rhobulge, xshift=-0.5cm, yshift=-3cm] {};
\draw[arrow] (vgas) -- (node9) |- (aN);
\draw[arrow] (vdisc) -- (node9) |- (aN);
\draw[arrow] (vbulge) -- (node9) |- (aN);
\draw[arrow] (vDM) -- (node9) |- (aN);

% connect baryonic v components to baryonic models
\draw [arrow] (vgas) -- (gasdisc);
\draw [arrow] (vdisc) -- (stellardisc);
\draw [arrow] (vbulge) -- (stellarbulge);

% connect DM halo to vDM
\draw [arrow] (DMhalo) -- (vDM);

% free parameter boxes
\node (fR0) [parameter, below of=vDM, xshift=-4cm, yshift=0cm, align=center] {$f_{R0}$};
\node (cvir) [parameter, right of=fR0, xshift=1cm, align=center] {$c_\mathrm{vir}$};
\node (vvir) [parameter, right of=cvir, xshift=1cm, align=center] {$v_\mathrm{vir}$};
\node (ML) [parameter, right of=vvir, xshift=1cm, align=center] {$\Upsilon$};
\node (sigma) [parameter, right of=ML, xshift=1cm, align=center] {$\sigma_g$};

% connect ML to stellar models
\node (node10) [junction, right of=vdisc, xshift=-1cm] {};
\draw [arrow] (ML) -- (node10) -- (stellardisc);
\draw [arrow] (ML) -- (node10) -- (stellarbulge);

% connect ML to aN
\draw[arrow] (ML) -- ($(vbulge)+(1cm,-0.5cm)$) -- (node9) |- (aN);

% connect vvir and cvir to DM halo
\draw [arrow] (cvir) -- (DMhalo);
\draw [arrow] (vvir) -- (DMhalo);

% connect fR0 to screening map and f(R) solver
\node (node11) [junction, left of=screeningmap] {};
\draw [arrow] (fR0) -- (node11) -- (screeningmap);
\draw [arrow] (fR0) -- (node11) -- (fRSolver);

% connect sigma to likelihood
\draw [arrow] (sigma) |- (node3) -- (node2) -- (likelihood);

% connect all free parameters to prior
\node (node12) [junction, right of=sigma, xshift=-1.5cm, yshift=1cm] {};
\draw [arrow] (sigma) -- (node12) |- (prior);
\node (node13) [junction, right of=ML, xshift=-1.5cm, yshift=1cm] {};
\draw [arrow] (ML) -- (node13) -- (node12) |- (prior);
\node (node14) [junction, above of=vvir, yshift=-1cm] {};
\draw [arrow] (vvir) -- (node14) -- (node13) -- (node12) |- (prior);
\node (node15) [junction, above of=cvir, yshift=-1cm] {};
\draw [arrow] (cvir) -- (node15) -- (node14) -- (node13) -- (node12) |- (prior);
\coordinate (p6) at ($(fR0)+(1.5cm, 1cm)$);
\draw [arrow] (fR0) -- (p6) -- (node15) -- (node14) -- (node13) -- (node12) |- (prior);

\end{tikzpicture}
    \caption{Pipeline illustrating the calculation of the posterior probability $\mathcal{P}(\theta | \mathcal{D})$ for parameter values $\theta$ and for a given galaxy with rotation curve data $\mathcal{D}$. As indicated in the legend, light green rectangles represent free parameters, ellipses represent tools or models, lozenges represent observational data, and dark green rectangles represent calculated quantities. The model illustrated here is an $f(R)$ model with varying $\fR$, a single mass-to-light ratio, an NFW halo and an added large-scale environment (i.e. Model C). Various details of the pipeline would differ in other models. For example, the nodes labelled `$\rho_\mathrm{env}$', `$\Phi_\mathrm{ext}$' and `screening map' would not be present in the case of Model B, which is identical to Model C but does not include a large-scale environment in the scalar field solver. Another example is Model F, which is identical to Model B but has two free parameters for the mass-to-light ratio, and would therefore also lose the environmental screening apparatus and further replace the node marked `$\Upsilon$' with two free parameter nodes: `$\Udisc$' and `$\Ubulge$'.}
    \label{F:Pipeline}
\end{figure*}
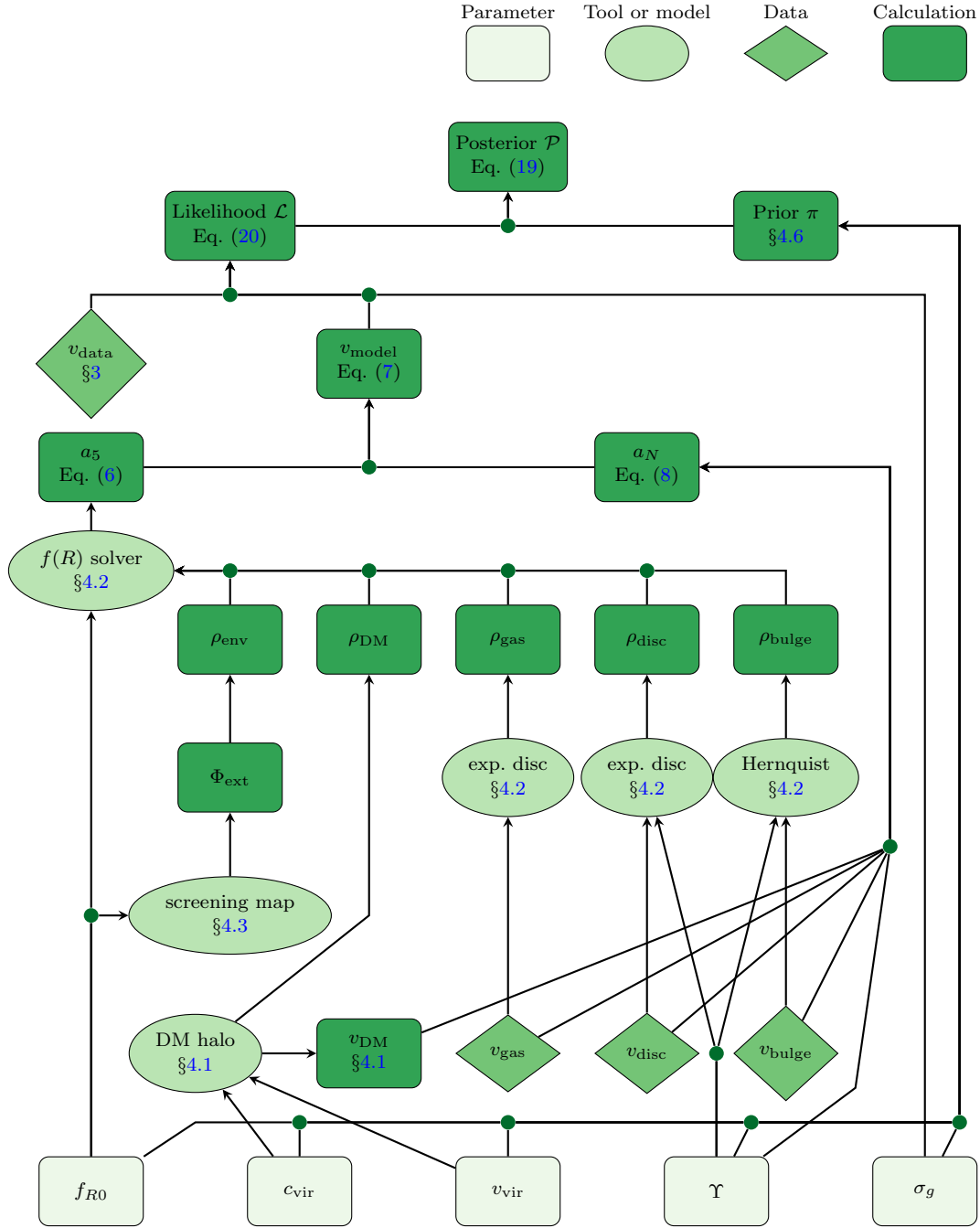

\subsection{Scalar Field Solver}
\label{S:Methods:fRSolver}

This subsection describes the calculation of the scalar field $f_R$ for a given mass distribution and cosmic background value $\fR$. The technique is essentially a 1D version of the Newton-Gauss-Seidel scalar field solver implemented in the $f(R)$ N-body code MG-GADGET \citep{Puchwein2013} which was used in Paper I.

Given the finding of Paper I that the scalar field profiles within galaxies typically adopt a discoid shape, the spherically-symmetric 1D approximation is not ideal but is made for reasons of computational cost; the jump to 2D corresponds to roughly two orders of magnitude in computational time, typically making it prohibitively expensive to perform a sufficient number of iterations to achieve MCMC convergence.

On the other hand, an analytic calculation of the screening radius and fifth force for a spherical body, such as that given in \citet{Sakstein2013}, would make the computation instantaneous. However, this prescription was found to be too inaccurate and sensitive to the (somewhat arbitrary) choice of outer limit of integration. Thus, the 1D numerical computation described in this subsection represents a reasonably accurate and reasonably fast compromise. The robustness of this 1D approximation is examined in detail in Appendix \S\ref{A:1D2D_comparison}, where a comparison to an axisymmetric 2D solver is presented.

To avoid unphysical positive values of the scalar field $f_R$ (e.g., due to finite step sizes in the iterative scheme), the calculation is performed in terms of the quantity $u \equiv \ln \frac{f_R}{\bar{f}_R\left(a\right)}$. Equation (\ref{E:FieldEOM}) can then be written as
\begin{equation}\label{E:EOM}
    \nabla^2e^u + \frac{1}{3c^2\bar{f}_R(a)}\left[\bar{R}\left(a\right)\left(1-e^{-\frac{u}{2}}\right) + 8\pi G \delta \rho\right] = 0.
\end{equation}

The role of the scalar field solver is to solve Equation~(\ref{E:EOM}) for $f_R$ across the galaxy, given an input density profile $\delta \rho$. This scalar field profile can then be used to calculate the fifth force contribution to the rotation curve via Eq. (\ref{E:a5}).

Equation~(\ref{E:EOM}) is discretised assuming spherical symmetry on a 1D radial grid. The outer edge of the grid needs to be at a radius larger than the Compton wavelength of the theory. For all of the $\fR$ values within the bounded prior (see \S\ref{S:Methods:Priors}), 5 Mpc is a sufficiently large value. Finer resolution is required at smaller radii than at these large radii, so it is appropriate to use logarithmically spaced grid cells, i.e. the radial gridlines are equally spaced in the coordinate $x \equiv \ln r$, with constant grid spacing $h_x$. It was found that 175 cells between $r_\mathrm{min}=0.05$kpc and $r_\mathrm{max}=5$Mpc gave sufficiently accurate, converged results.

Defining
\begin{equation}\label{E:Lij}
    \mathcal{L}_{i} \equiv \left(\nabla^2e^u\right)_{i} +  \frac{1}{3c^2\bar{f}_R(a)}\left[\bar{R}\left(a\right)\left(1-e^{-\frac{u_{i}}{2}}\right) + 8\pi G \delta \rho_{i} \right],
\end{equation}
where the index $i$ denotes the radial grid cells, Equation (\ref{E:EOM}) is then
\begin{equation}\label{E:DiscreteEOM}
    \mathcal{L}_{i}=0.
\end{equation}
This is solved, as in MG-GADGET, with an iterative Newton-Gauss-Seidel approach, where at iteration $n$ the scalar field is updated via
\begin{equation}\label{E:NGSUpdate}
    u_{i}^{n+1} = u_{i}^{n} - \frac{\mathcal{L}_{i}^n}{\frac{\partial\mathcal{L}_{i}^n}{\partial u_{i}^n}}.
\end{equation}
In order to do this, we need discretised expressions for the Laplace operator on our grid, as well as the quantity $\frac{\partial\mathcal{L}_{i}^n}{\partial u_{i}^n}$. The Laplace operator in the coordinate $x \equiv \ln r$ is given by
\begin{equation}\label{E:Laplace}
    \nabla^2f = \frac{1}{r^3}\frac{\partial}{\partial x}\left(r\frac{\partial f}{\partial x}\right)
\end{equation}
which is discretised as
\begin{equation}
    \left(\nabla^2f\right)_{i} = \frac{1}{r_i^3 h_x^2} \left( r_{i+\frac{1}{2}}(f_{i+1} - f_{i}) - r_{i-\frac{1}{2}}(f_{i} -f_{i-1})\right),
\end{equation}
where $r_i$ indicates the radial position of the cell centre of cell $i$, while $r_{i-\frac{1}{2}}$ and $r_{i+\frac{1}{2}}$ are the positions of the inner and outer cell boundaries. Finally, the quantity $\frac{\partial\mathcal{L}_{i}}{\partial u_{i}}$ is given by
\begin{equation}
    \frac{\partial\mathcal{L}_{i}}{\partial u_{i}} = \frac{\bar{R}(a)}{6c^2\bar{f}_R(a)}e^{-\frac{u_{i}}{2}} - e^{u_{i}}\left(\frac{r_{i+\frac{1}{2}} + r_{i-\frac{1}{2}}}{r_i^3 h_x^2} \right).
\end{equation}

Iterations of Eq. (\ref{E:NGSUpdate}) are performed until $\Delta u_i \equiv u_{i}^{n+1} - u_{i}^{n}$ is less than $10^{-7}$ at all grid cells $i$. Through experimentation, this tolerance level has been found to give sufficiently accurate, converged results.

All that remains now is to provide a density profile $\rho_i$ for Eq. (\ref{E:Lij}). For a given galaxy, the density profile depends on the parameter choices for the mass-to-light ratio(s) and the two dark matter halo parameters. Density data for the baryonic components are not provided with the SPARC data, so we have instead fitted $v_\mathrm{gas}$ and $v_\mathrm{disc}$ with exponential disc profiles and $v_\mathrm{bulge}$ with a Hernquist profile. In the cases of the 2D exponential discs, we spherically average the profiles for use in the 1D solver. These density profiles are then fed to the scalar field solver (multiplied by the mass-to-light ratio in the case of the stellar disc and bulge), along with the spherical dark matter halo profile, either NFW or DC14.

\subsection{Environmental Screening}
\label{S:Methods:EnvScreening}

The environment of an object plays a role in its scalar field profile. As verified in Paper I, a galaxy embedded within a large-scale overdensity experiences an effective $\fR$ that is lower than the cosmic background value, and as a result will have a larger screening radius than that of the case in which the galaxy occupied a region of cosmic mean density.

It has been shown in simulations \citep{Zhao2011a, Zhao2011b, Cabre2012} that as a first approximation, the degree of environmental screening of a given galaxy can be quantified by the gravitational potential due to external sources $\Phi_\mathrm{ext}$.

For this very purpose of measuring the impact of environmental screening on tests of gravity, two groups---\mbox{\citet{Cabre2012}} and \citet{Desmond2018c}---have constructed `screening maps': 3D maps of $\Phi_\mathrm{ext}$ throughout the local universe. In this work, we use the screening map of \citet{Desmond2018c} which uses updated techniques and catalogues compared to that of \citet{Cabre2012}. Full details regarding the construction of the map can be found in the original paper, but it is worth noting that the value of $\Phi_\mathrm{ext}$ given by the screening map at a given point in space depends on the adopted value for $\fR$, because the map sums contributions from all mass within the Compton wavelength $\lambda_C$ of the considered point, which relates to $\fR$ via the approximate equation \citep{Cabre2012}
\begin{equation}\label{E:Compton}
\lambda_C \approx 32 \sqrt{\frac{|\fR|}{10^{-4}}} \ \mathrm{Mpc}.
\end{equation}

In this work, we use the screening map of \citet{Desmond2018c} to find $\Phi_\mathrm{ext}$ for each of the 147 galaxies that remain in the sample after making the first four cuts described in \S\ref{S:Data}. The screening map uses a Bayesian methodology to reconstruct full posteriors on $\Phi_\mathrm{ext}$ at any point in space within $\sim 200$Mpc, by propagating uncertainties on the various inputs. For each of these galaxies, the posterior distribution for $\Phi_\mathrm{ext}$ is calculated for a log-spaced sequence of 20 values of $\fR$, the same values used in Models H0-19 (see Table \ref{T:HVals}). This information is then used in two ways.

Firstly, we exclude from the sample any galaxy for which the  $1 \sigma$ upper bound on $|\Phi_\mathrm{ext}|/c^2$ is greater than $\frac{3}{2} \afR$ (which is an approximate criterion for environmental screening, see \citealt{Cabre2012}), for \textit{any} of the 20 considered values of $\fR$. This is tantamount to removing from the sample any galaxies for which we believe environmental screening would be a significant effect in an $f(R)$ Universe. As indicated in \S\ref{S:Data}, this cuts 62 of the 147 remaining galaxies once the other sample cuts have already been performed, leaving our final sample of 85 galaxies.

Secondly, we investigate the effect of the environmental contribution to the scalar field profile of the remaining galaxies in our sample. We consider a spherical top hat model, with a radius equal to the Compton wavelength $\lambda_C$, and a gravitational potential equal to the maximum posterior value of $\Phi_\mathrm{ext}$. As indicated by the quantity $\rho_\mathrm{env}$ in Figure \ref{F:Pipeline}, this structure is added to the overall density profile before it is fed to the scalar field solver. As shown in Table \ref{T:Models}, this change is only made in the case of Model C, which is otherwise identical to Model B, i.e. $f(R)$+varying $\fR$+NFW. The results of Model C can be compared to those of Model B to check the robustness of our results against the residual effects of environmental screening in our final sample. Note that because $\Phi_\mathrm{ext}$ is a function of $\fR$, and $\fR$ is a free parameter in Model C, the mass and size of this large-scale structure needs to be recalculated with each Monte Carlo realisation.

\begin{figure}
    \centering
    \includegraphics{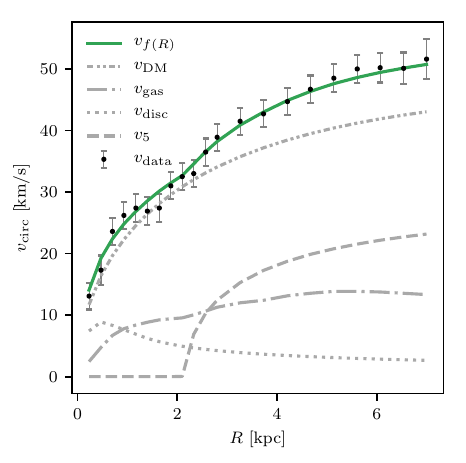}
    \caption{Shown as an example, the rotation curve of NGC3741 and corresponding fit. The black points with errorbars show the observed rotation curve, while the green curve gives the $f(R)$ (Model B) fit. The grey textured lines show the contributions to the $f(R)$ fit of the various components: gas, stellar disc, dark matter, and fifth force, as labelled. The galaxy does not have a bulge component, so there is no $v_\mathrm{bulge}$ shown. It should be noted that this fit is unrepresentatively good; it was chosen because there is a clear upturn-like feature that is well fit with a screening radius and fifth force. As is shown in later figures, not all galaxies are fit that well by the model.}
    \label{F:FitExample}
\end{figure}

\subsection{Stellar Self-screening}
\label{S:Methods:StellarScreening}

An assumption underlying our analysis is that the scalar field $f_R$ is sourced by all of the mass components: the dark matter, gas, and stars. However, in an $f(R)$ universe, one might expect a significant population of stars to be `self-screened', and thus neither source nor couple to the fifth force \citep{Davis2012}.

This effect was the underlying principle for the aforementioned works by \citet{Vikram2018} and \citet{Desmond2019, Desmond2018a, Desmond2018b}, which all searched for signals predicted to result from this differential coupling of the fifth force to stars and gas.

For the purposes of the present work, the expected effect of the stellar self-screening is that for a given $\fR$, the screening radius will be smaller than if the effect is ignored. This was demonstrated in Paper I.

The contribution of a given star depends on its mass, radius, and environment. Thus, performing a fully self-consistent calculation incorporating the detailed contribution of the entire stellar population is difficult. We instead repeat the approach adopted in Paper I: we implement the extreme scenario in which all stars are assumed to be fully self-screened, and not sourcing the scalar field at all. In practice, this simply amounts to omitting the stellar input to the scalar field solver.

As in the test for environmental screening in the previous subsection, we employ this technique only for one model, Model D, which is otherwise identical to Model B, i.e. $f(R)$+varying $\fR$+NFW. The `true' result should then be bookended by these two extremes, so the results of Model D can be compared to those of Model B in order to gain an undestanding of the error induced in our inference by the effects of stellar self-screening.

\subsection{Markov Chain Monte Carlo (MCMC)}
\label{S:Methods:MCMC}

We use an affine-invariant, parallel-tempered Markov chain Monte Carlo (MCMC) technique, using the publicly available \textsc{python} package \textsc{emcee} \citep{Foreman2013} to explore the posteriors of the free parameters of each model.

For a given galaxy with rotation curve data $\mathcal{D}$, the posterior probability $\mathcal{P}(\theta | \mathcal{D})$ for a given set of parameters $\theta$ is given by
\begin{equation}
\label{E:Posterior}
    \mathcal{P}(\theta|\mathcal{D}) \propto \mathcal{L}(D |\theta)\pi(\theta),
\end{equation}
where $\mathcal{L}(D|\theta)$ represents the likelihood function, and $\pi(\theta)$ represents the prior probability distributions for the parameters $\theta$. The choices of priors will be discussed further in \S\ref{S:Methods:Priors}.

The pipeline for calculating this posterior probablity, for a given galaxy and a set of parameters $\theta$, is illustrated heuristically in Figure \ref{F:Pipeline}. The diagram specifically illustrates the pipeline for an $f(R)$ model with freely varying $\fR$ and an environmental contribution (i.e. Model C); some of the details of the diagram would change for other models, and examples of such changes are given in the accompanying caption.

Assuming that the errors $\sigma$ on the data are Gaussian, the log-likelihood function $\ln \mathcal{L}(\mathcal{D}| \theta)$ for a given SPARC galaxy is given by
\begin{equation}\label{E:Likelihood}
    \ln \mathcal{L} = -\frac{1}{2}\sum_{j}\left[ \left(\frac{ v_{\mathrm{data}_j} - v_\mathrm{model}(r_j)}{\sigma_j}\right)^2 + \ln\left(2\pi\sigma_j^2\right) \right],
\end{equation}
where the sum is over the data points of an individual rotation curve.

In addition to the observational errors provided with the data points, for each galaxy we add in quadrature an additional constant error component $\sigma_g$, i.e.,
\begin{equation}\label{E:Errors}
    \sigma_j^2 = \sigma_{\mathrm{obs}, j}^2 + \sigma_{g}^2.
\end{equation}
$\sigma_g$ is then left as a free parameter in the fit for each galaxy. This additional error term is included in order to account for galaxy features not included in the model, e.g. spiral arms and other baryonic features, but a fit is penalised for adopting too large a value of $\sigma_g$, via the second term in the log-likelihood Eq. (\ref{E:Likelihood}).

For each MCMC run, we use 30 walkers, 4 temperatures, and 5000 iterations (after burn-in). The chains have all been checked for convergence, both visually and by ensuring that the Gelman-Rubin statistic \citep{Gelman1992} $\mathcal{R}$ is sufficiently close to 1. In particular, for the overwhelming majority of the fits, $|\mathcal{R}-1| < 0.01$, apart from one or two galaxies in some models, which exhibit some bimodality.

An example of a fit generated with this method is shown in Figure \ref{F:FitExample}. The figure shows the observed rotation curve of NGC3741, along with the best fit (i.e. maximum a posteriori) rotation curve under the $f(R)$ model with freely varying $\fR$ (i.e. Model B). Also shown are the various components: $v_\mathrm{disc}$, $v_\mathrm{gas}$, $v_\mathrm{DM}$, which combine in quadrature via Eq. (\ref{E:v_model}) to give the model.

A fourth component is shown in Figure \ref{F:FitExample}: $v_5$. This is the contribution of the fifth force to the rotation curve model, given by
\begin{equation}
    v_5(r) = \sqrt{a_5(r) r}
\end{equation}

The only contribution to the model that is not a function of fit parameters is the gas curve $v_\mathrm{gas}$, which is provided with the data as a fixed quantity. $\vdisc$ is also provided with the data, but its overall normalisation is set by the mass-to-light ratio $\Upsilon$. It can be seen in the $v_5$ curve that the fit places a screening radius at around 2 kpc. This results in a mild upturn in the model rotation curve, which visibly appears to give a good fit to the data. It should be noted that this rotation curve was chosen because of this ideal behaviour, and is not necessarily representative of the sample at large.

\subsection{Priors}
\label{S:Methods:Priors}

\begin{figure}
    \centering
    \includegraphics{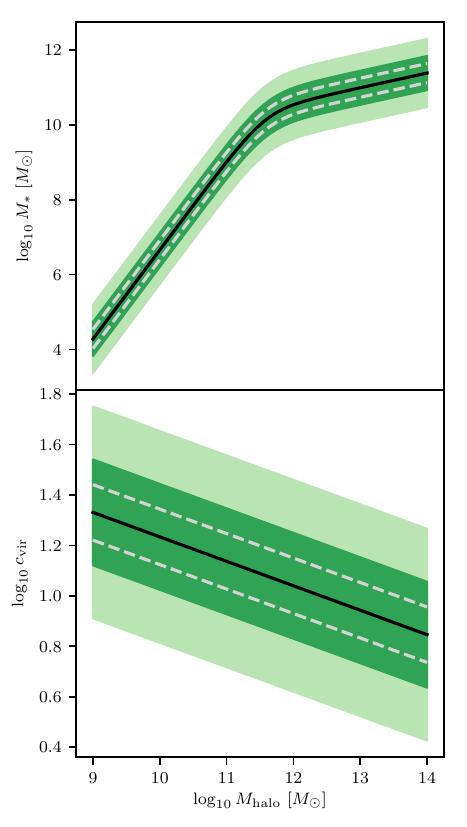}
    \caption{\textit{Top:} Stellar mass-halo mass relation based on \citet{Moster2013} used as a log-normal prior. The dark and light green regions indicate the $1$ and $2\sigma$ regions after broadening the relation to account for $f(R)$ gravity effects (see text for details) respectively. The light grey dashed lines indicate the $1\sigma$ region before the broadening. \textit{Bottom:} As with the top panel, but for the concentration-halo mass relation from \citet{Dutton2014}.}
    \label{F:Priors}
\end{figure}

As indicated in Table \ref{T:Models}, different models have different free parameters. For the models with a freely varying $\fR$, we adopt a flat prior in $\log_{10}\afR$, between $|\fR|=10^{-9}$ and $|\fR|=2\times 10^{-6}$. For even larger values of $|\fR|$ the model would struggle to screen the Milky Way at the solar radius, while even smaller values would be of limited interest due to the minuteness of the deviations from $\Lambda$CDM. 

For any given galaxy, we adopt a flat prior for $\sigma_g$, between 0 and twice the maximum observed error for that galaxy.

The priors for the remaining parameters---the two dark matter parameters, $c_\mathrm{vir}$ and $v_\mathrm{vir}$, and where applicable, the one or two mass-to-light ratios $\Upsilon$---are less agnostic. Using flat priors was found to have the result that the fits are able to artificially inflate or deflate their halo masses so as to have a screening radius within the radial range of the rotation curve. As a consequence, unphysical vertical clustering features would appear in the stellar mass-halo mass diagrams and concentration-halo mass relation diagrams of the best fit models.

To avoid this undesirable behaviour, we follow the approach of \citet{Katz2017} and use empirical stellar mass-halo mass and concentration-halo mass relations from \citet{Moster2013} and \citet{Dutton2014} respectively as lognormal priors. These relations are depicted in Figure \ref{F:Priors}. For a fixed mass-to-light ratio, these relations translate directly to priors on $v_\mathrm{vir}$ and $c_\mathrm{vir}$ respectively. In the case of a freely varying $\Upsilon$, there is an increased freedom, as the stellar mass depends on the value of $\Upsilon$.

It should be noted that these relations are derived from $\Lambda$CDM simulations, so their applicability in an $f(R)$ universe is not entirely clear. $f(R)$ gravity simulations, such as those of \citet{Mitchell2019} suggest that for $\afR=10^{-6}$ (denoted F6) and below, halo concentrations are enhanced by 0.1 dex or so. Taking this into account, we increase the width of our adopted concentration-halo mass relation by 0.1 dex; approximately equivalent to a doubling of the width. The priors should then encompass the `true' relations of both a $\Lambda$CDM and an $f(R)$ universe.

For the stellar mass-halo mass relation, galaxy formation simulations are not yet available in the $f(R)$ paradigm, so the $f(R)$ relation is unknown. However, the work of e.g. \citet{Cataneo2016} suggests that the maximum increase to the halo mass function at the F6 level is small; approximately 10\% or so. Assuming baryonic feedback mechanisms to have largely similar effects in an $f(R)$ universe, one might then expect the stellar mass-halo mass relation to be changed by a similar degree. Thus, a doubling of the width of the relation should also comfortably encompass the `true' relation in the $f(R)$ paradigm. We approximately achieve this by adding 0.2 dex to the width.

As a final note, we follow \citet{Katz2017} in imposing some additional constraints. Firstly, we require that the baryon mass fraction is always less than the cosmological value. Furthermore, the parameters $\log_{10} c_\mathrm{vir}$, $\log_{10} \frac{v_\mathrm{vir}}{1 \mathrm{m/s}}$, and $\log_{10}\frac{\Upsilon}{1 M_\odot/L_\odot}$ are constrained to lie within the bounds (0, 2), (4, 5.7), and (-0.52, -0.1) respectively. The constraints on $c_\mathrm{vir}$ and $v_\mathrm{vir}$ are rather loose, while the constraint on $\Upsilon$ is somewhat tighter, but informed by the stellar synthesis models of \citet{Meidt2014, McGaugh2014, Schombert2014}.

% RESULTS
\section{Results}
\label{S:Results}

\subsection{Constraints on $\fR$}
\label{S:Results:fR0}

\begin{figure}
    \centering
    \includegraphics{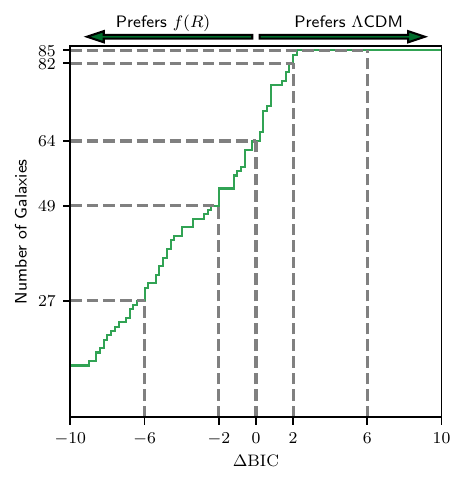}
    \caption{Cumulative distribution function of $\Delta\mathrm{BIC}=\mathrm{BIC}_{f(R)}-\mathrm{BIC}_{\Lambda\mathrm{CDM}}$ across SPARC galaxies (total=85), where BIC is the Bayesian Information Criterion, given by Eq.~(\ref{E:BIC}). The $f(R)$ model here is Model B (freely varying $\fR$), while the $\Lambda$CDM model is Model A, both of which use NFW haloes and single mass-to-light ratios. As indicated by the arrows at the top of the Figure, a negative value for $\Delta$BIC indicates a preference for $f(R)$, and a positive value for $\Lambda$CDM. The grey dashed lines mark the fiducial values of $|\Delta\mathrm{BIC}|=0, 2, 6$. Corresponding numbers on the $y$-axis indicate numbers of galaxies \textit{below} the lines. For example, 49 galaxies have a mildly significant preference for $f(R)$, while $85-82=3$ galaxies have a similar preference for $\Lambda$CDM. This figure indicates that the majority of galaxies prefer \textit{some} $f(R)$ model, but it is worth noting that this does not necessarily mean that they all prefer the same $\fR$ value.}
    \label{F:dBIC_CDF} 
\end{figure}

\begin{figure}
    \centering
    \includegraphics{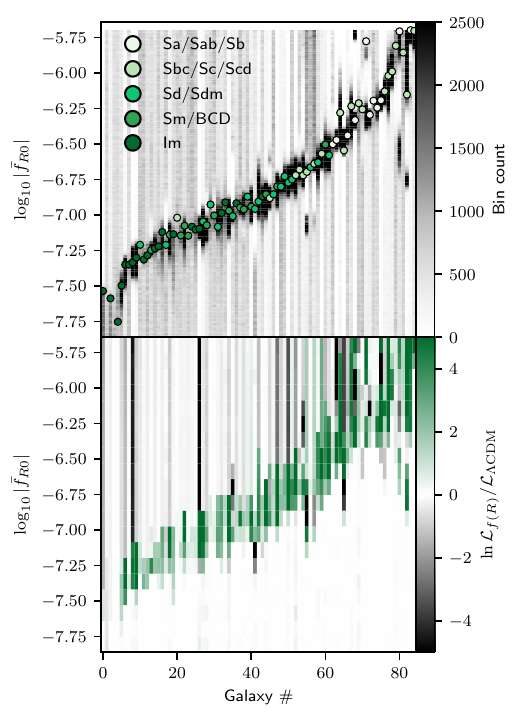}
    \caption{\textit{Top:} For all 85 galaxies in the sample, the marginal posterior distributions for $\fR$ under the $f(R)$ model with freely varying $\fR$ and NFW haloes (i.e. Model B) are shown. Each individual posterior is a histogram with 200 bins, equally log-spaced between the edges of the $\fR$ prior. The 85 histograms are then vertically juxtaposed, and ordered by the position of the peak of the histogram. The colourscale is truncated at a bin count of 2500 to allow sufficient contrast in less favoured regions, but actually reaches up to $\sim$ 30000 in some cases. The points show the $\fR$ values of the best fit models, coloured according to the Hubble classification of the galaxy, as indicated in the legend. The marginal posteriors do not appear to show any significant clustering around a single value for $\fR$. \textit{Bottom:} For a series of 20 $f(R)$ models with fixed $\fR$ (i.e. Models H0-19), the likelihood ratios $\ln\mathcal{L}_{f(R)}/\mathcal{L}_{\Lambda\mathrm{CDM}}$, where the $\Lambda$CDM model used is Model A, and likelihoods are calculated for the best fit models. This colourscale is also truncated, at a likelihood ratio of $\pm 5$. This panel conveys similar information to the top panel: different galaxies appear to prefer different ranges of $\fR$. However, as discussed in the text, the possibility remains of a single global value that is consistent with all galaxies, particularly in the region $\afR \sim 10^{-7}$, where many of the galaxies are fully screened and therefore insensitive.}
    \label{F:Posteriors}
\end{figure}

\begin{figure*}
    \includegraphics{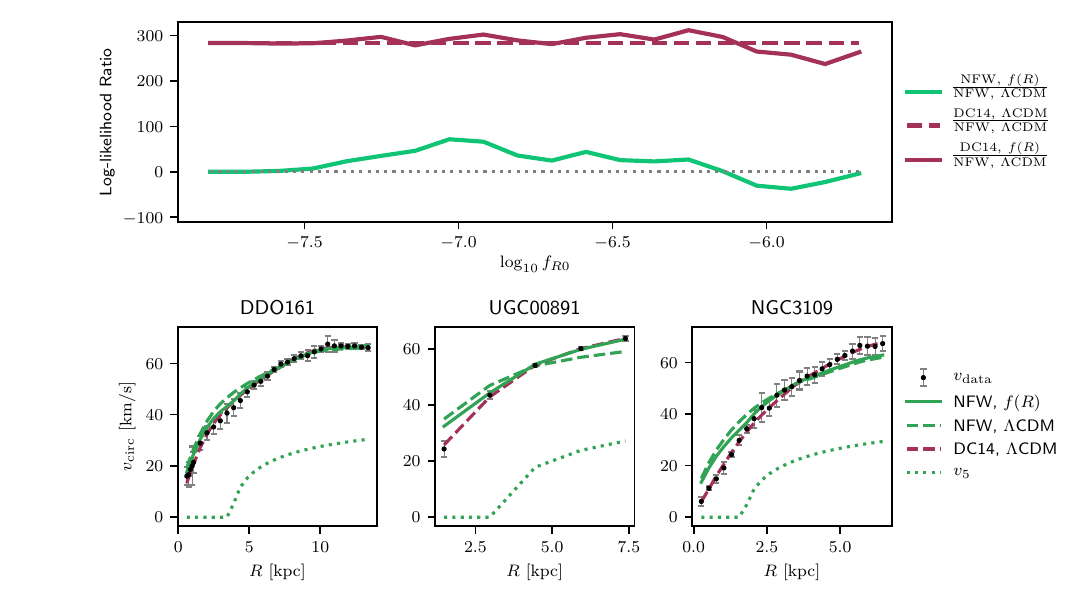}
    \centering
    \caption{\textit{Top:} The coloured lines show model comparisons via the log-likelihood ratio $\ln\mathcal{L}_1/\mathcal{L}_2$. The dashed purple line shows the log-likelihood ratio for DC14+$\Lambda$CDM (Model G) versus NFW+$\Lambda$CDM (Model A). The green solid line shows the log-likehood ratios for a range of fixed-$\fR$ models with NFW haloes (Model H) versus NFW+$\Lambda$CDM (Model A), while the purple solid line shows the ratios for another range of fixed-$\fR$ models, this time with DC14 haloes (Model I), again versus Model A. The grey dashed line shows where the log-ratio equals zero, i.e. where both models are equally favoured. \textit{Bottom:} Rotation curves and fits of three galaxies: DDO161 (\textit{left}), UGC00891 (\textit{centre}), and NGC3109 (\textit{right}). In each case, the observed rotation curve is shown, as well as 3 fits: Model A (green dashed line), Model G (purple dashed), and Model H7 (green solid), i.e. the $f(R)$+NFW model imposing $\log_{10}\afR=-7.03$, corresponding to the peak of the green likelihood ratio curve in the top panel. The three galaxies chosen are those which contribute the most to this peak. This figure shows the key result of the paper: the galaxies which most favour $f(R)$ over $\Lambda$CDM are those for which NFW provides a poor fit, and using a cored halo profile under $\Lambda$CDM gives a better fit to the rotation curves than cuspy haloes with $f(R)$ gravity. Furthermore, when cored haloes are assumed, $f(R)$ gravity does not give any significant improvement in the agreement with the data, as can be seen by comparing the purple solid and purple dashed curve in the top panel.}
    \label{F:lnLRatios}
\end{figure*}

A widely used method for model comparison is the Bayesian Information Criterion \citep{Schwarz1978}, given by
\begin{equation}
\label{E:BIC}
    \mathrm{BIC} = \ln(n)k - 2\ln(\mathcal{L}),
\end{equation}
where $n$ is the number of data points, $k$ is the number of parameters of the model, and $\mathcal{L}$ is the maximised likelihood of the model. The first term acts on behalf of Occam's razor, penalising overcomplicated models with too many parameters. Two models, 1 and 2, can then be compared by calculating the difference in the BIC: $\Delta\mathrm{BIC}\equiv \mathrm{BIC}_1 - \mathrm{BIC}_2$. A positive (negative) value for this quantity indicates a preference for model 2 (1). We define values in the range $|\Delta\mathrm{BIC}|>2$ as `mildly significant', and values in the range $|\Delta\mathrm{BIC}|>6$ as `strongly significant' \citep{Kass1995}.

Figure \ref{F:dBIC_CDF} shows a first comparison of an $f(R)$ model with freely varying $\fR$ and a $\Lambda$CDM model, in both cases using NFW haloes (i.e. Models B and A respectively). The cumulative distribution function of $\Delta\mathrm{BIC}\equiv \mathrm{BIC}_B - \mathrm{BIC}_A$ is displayed. Model B treats $\fR$ as a free parameter and so has one more free parameter than Model A, for which it is penalised by the first term in Eq. (\ref{E:BIC}). Nonetheless, the majority of galaxies show \textit{some} preference for $f(R)$, with 64/85 galaxies having $\Delta\mathrm{BIC}<0$. Perhaps more striking are the numbers of galaxies that have mildly significant ($|\Delta\mathrm{BIC}|>2$) preferences either way, with 3 galaxies preferring $\Lambda$CDM and 49 galaxies preferring $f(R)$. No galaxies have a strongly significant ($|\Delta\mathrm{BIC}|>6$) preference for $\Lambda$CDM, while 27 galaxies do for $f(R)$.

It is worth noting that the galaxies with the strongest preferences for $f(R)$ also typically infer much lower values for $\sigma_g$ under Model B than under Model A; Model A tries to compensate for a poor fit by increasing the scatter.

These numbers, and the preference for $f(R)$ implied by them, is perhaps unsurprising. Given a freely varying $\fR$, the fits are able to place a screening radius anywhere within the radial range of the rotation curve, and are thus able to choose a favourable position such that the fit is improved. It is only in a minority of cases that no such position can be found. For these galaxies, a low $\afR$ is chosen such that the galaxy is fully screened. This results in $\mathcal{L}_{f(R)}/\mathcal{L}_{\Lambda\mathrm{CDM}}\rightarrow 1$ and thus $\Delta\mathrm{BIC} \rightarrow \ln(n)$. For the SPARC galaxies, this will typically fall in the range $2-5$, and so it is impossible for a galaxy to have a `strongly significant' preference for $\Lambda$CDM over an $f(R)$ model with a freely varying $\fR$. This just reflects the fact that $\Lambda\mathrm{CDM}$ is contained within the $f(R)$ model as a limiting case.

\begin{figure*}
    \includegraphics{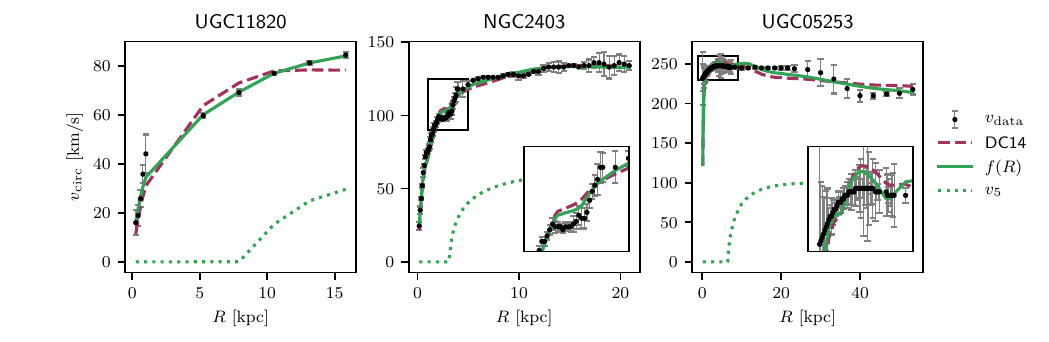}
    \caption{Rotation curves and fits of UGC11820 (\textit{left}), NGC2403 (\textit{middle}), and UGC05253 (\textit{right}). These are the three galaxies that have the strongest preference for an $f(R)$ model with an NFW halo and a freely varying $\fR$ (i.e. Model B) over a $\Lambda$CDM model with a cored DC14 halo (i.e. Model G). The Model B and G fits are consequently the ones shown in the panels, with green solid and purple dashed lines respectively. To give an indication of where the $f(R)$ fit is placing the screening radius, each panel also shows the contribution to the $f(R)$ fit of the fifth force, i.e. $v_5$, plotted as a green dotted line. The inset panels for NGC2403 and UGC05253 give magnified views of crowded areas of the rotation curves. UGC11820 is a success for $f(R)$, which gives a markedly better fit than the $\Lambda$CDM+DC14 model, both visually and formally. In the other two cases, however, neither model does particularly well. From left to right, the best-fit values of $\log_{10}\afR$ are -7.07, -6.57, and -6.19.}
    \label{F:BvG}
\end{figure*}

Of course, in an $f(R)$ Universe, $\fR$ would not vary freely from galaxy to galaxy as we have allowed it to here. Instead, there would be a single global value for $\fR$. The natural question then is whether the individually inferred values for $\fR$ show any clustering around a single global value. The upper panel of Figure \ref{F:Posteriors} addresses this question, showing the marginal posterior distributions for $\fR$ across all the galaxies, fitting again with Model B (freely varying $\fR$+NFW). The overlaid coloured points show the values of $\fR$ for the best fit models, which typically coincide with the peaks of the marginal posteriors, except in cases where there is some multimodality or $\fR$ is not well constrained. 

The key result shown in this panel is that there does not appear to be any immediate sign of clustering around a single value of $\fR$. For most galaxies, $\fR$ values in a reasonably narrow range are preferred, but these inferred values are almost evenly spread across two orders of magnitude, which is inconsistent with a single global value.

As an additional note, the Hubble types of the galaxies are indicated in the upper panel of Figure \ref{F:Posteriors}. The classifications used are those that accompany the SPARC data, and match the classification scheme of \citet{deVaucouleurs1991}: Sa-Sd types are spiral galaxies with spiral arms of decreasing tightness, Sm and Im types are Magellanic spirals and irregulars, and BCD types are `blue compact dwarfs'. There appears to be a correlation between inferred $\fR$ and Hubble type. This is to be expected, as dwarf galaxies would be sensitive to lower values of $\afR$, for which larger galaxies would be entirely screened. Conversely, larger galaxies would be sensitive to higher values, for which dwarf galaxies would be fully unscreened. 

This finding demonstrates an important caveat to the above result that there is no obvious global value for $\fR$: different galaxies have different sensitivity ranges for $\fR$, so there would naturally be a spread in fitted $\fR$ values when comparing across the whole sample. Even in an $f(R)$ Universe, only the galaxies with the true $\fR$ within their sensitivity ranges would fit this value, and the other galaxies would instead just choose the incorrect $\fR$ values that most improved their fits. 

It is important then to invert the question; rather than asking whether all galaxies are consistent with one $\fR$ value, we can ask whether one $\fR$ value is consistent with all galaxies, i.e. whether a specific $\fR$ value can be imposed globally that gives rotation curve models consistent with the observations of all galaxies, even if it is not the most preferred $\fR$ value for a significant subpopulation of galaxies, possibly lying outside of their sensitivity range.

To address this question, we consider a series of 20 $f(R)$ models with globally imposed $\fR$, i.e. Models H0-19. The $\fR$ values utilised are given in Table \ref{T:HVals}. For each of the 20 $\fR$ values and each of the 85 galaxies, the lower panel of Figure \ref{F:Posteriors} shows the likelihood ratio $\ln\mathcal{L}_{f(R)}/\mathcal{L}_{\Lambda\mathrm{CDM}}$, taking the best fit parameters under Models H$x$ and A. This panel conveys similar information to the top panel; visually, there does not appear to be any obvious global value for $\fR$. However, there is additional information to be gleaned. In the higher regions, $\log_{10}\afR \gtrsim -6.5$, there are several galaxies which prefer such values of $\fR$. However, there are also a great many galaxies for which these values provide a worsening of the fit compared with $\Lambda$CDM. On the other hand, in the region $\log_{10}\afR \sim -7$, there are several galaxies which prefer such values, and the galaxies which prefer higher values are indifferent towards it, as a result of being completely screened in this region of $\fR$. This region, therefore, appears to be the most promising within our sensitivity range.

A side-note is that the lower panel of Figure \ref{F:Posteriors} also gives an understanding of the sensitivity of the sample. Beneath the diagonal green region is a white region, where galaxies are screened and $\Lambda$CDM and $f(R)$ are indistinguishable. As we go to progressively lower values of $\afR$, more galaxies inhabit the white region. We define $\log_{10}\afR \sim -7.2$ or $\afR \sim 6 \times 10^{-8}$ as the `sensitivity level' of our sample, being the point at which roughly the half of the sample has become insensitive. 

Rather than considering each galaxy individually, we can perform a model comparison across the sample as a whole. For each of the Models H0-19, we can perform an overall model comparison with $\Lambda$CDM by calculating the total likelihood ratio $\ln \mathcal{L}_{f(R)}/\mathcal{L}_{\Lambda\mathrm{CDM}}$. The log-likelihoods $\ln\mathcal{L}$ are calculated with Eq. (\ref{E:Likelihood}) taking the best fit (i.e. maximum a posteriori) parameters for each galaxy, then summing the contributions of all galaxies in our sample. A more fully Bayesian approach would be to marginalise over the full parameter space rather than taking the best fit parameters. The likelihoods then become \textit{marginal} likelihoods, and their ratios are known as Bayes factors. The calculation of these Bayes factors, however, is notoriously expensive so we instead consider the more easily calculable likelihood ratios, which are nonetheless robust tools for model comparison. The green curve in Figure \ref{F:lnLRatios} shows the result of this likelihood ratio calculation for the 20 $\fR$ values listed in Table \ref{T:HVals}.

Figure \ref{F:lnLRatios} shows that $f(R)$ is disfavoured for values $\log_{10}\afR \gtrsim -6.1$, reaching a decrease in log-likelihood of $\sim 30$ at its most extreme. Note that at higher values still, the likelihood for $f(R)$ starts to approach that of $\Lambda$CDM again. However, this corresponds to the extreme and unphysical case in which every galaxy is fully unscreened, and so the rotation curve models exactly resemble those of $\Lambda$CDM, but with an overall mass reduced to $3/4$ of the inferred $\Lambda$CDM mass, to account for the gravitational accelerations being $4/3$ times that of standard gravity. 

At the opposite end of the spectrum, $\log_{10}\afR \lesssim -7.6$, the likelihood ratios again approach identity. This corresponds to the regime in which every galaxy is fully screened, so the fits are now exactly identical to those of $\Lambda$CDM.

At intermediate values of $\fR$, $f(R)$ appears to be favourable compared to $\Lambda$CDM. The peak of the signal is for the model with $\log_{10}\afR = -7.03$, where there is a rather significant increase in the log-likelihood of around 66.

The lower panels of Figure \ref{F:lnLRatios} show the rotation curves and fits of the three galaxies with the highest individual likelihood ratios for Model H7 ($\log_{10}\afR=-7.03$) compared with Model A. In other words, the three galaxies that contribute the most to the peak in the upper panel of Figure \ref{F:lnLRatios}. These galaxies are DDO161 ($\ln \mathcal{L}_{f(R)}/\mathcal{L}_{\Lambda\mathrm{CDM}}=14.96$), UGC00891 (6.14), and NGC3109 (6.13). 

It is striking that the three galaxies are all rather similar; dwarf galaxies (Hubble types all Im or Sm) with similar radial extents and maximum rotation speeds. Furthermore, in each case, the behaviour of the fits is similar. The observed rotation curves are poorly fit with cuspy NFW profiles, which place too much mass in the inner regions in order to fit the velocity data in the outer regions. The reason the $f(R)$ fits provide a significant improvement is that the presence of the fifth force in the outer regions allows a reduced mass in the inner regions.

The DC14 profile for dark matter haloes \citet{DiCintio2014} is empirically derived from $\Lambda$CDM simulations incorporating stellar feedback. The inner slope of the profile depends on the stellar content of the galaxy, giving a more cored profile to galaxies with intermediate ($\sim 0.5 \,  \%$) stellar mass fractions. \citet{Katz2017} found that fitting the SPARC galaxies with a DC14 profile gives a consistent improvement compared to NFW, particularly in cases such as the galaxies depicted in the lower panels of Figure \ref{F:lnLRatios}, i.e. dwarf galaxies requiring a cored profile.

The upper panel of Figure \ref{F:lnLRatios} also shows a comparison of Models G and A (the dashed purple line), i.e. a model comparison of $\Lambda$CDM models with DC14 and NFW haloes. The increase in log-likelihood is nearly 300; a very significant improvement, and far more significant than the increase in log-likelihood for any given $f(R)$ model with an NFW halo. The reason for this improvement is readily apparent in the rotation curve fits in the lower panels of Figure \ref{F:lnLRatios}. Echoing the findings of \citet{Katz2017}, the DC14 fits to the three galaxies give consistently better agreement with the data than NFW. We find that this holds for both $\Lambda$CDM and $f(R)$ gravity.

The final model comparison shown in the upper panel of Figure \ref{F:lnLRatios} is between Models I0-19 and Model A, i.e. a comparison of DC14+$f(R)$ and NFW+$\Lambda$CDM. Model I does not give a marked improvement to the fits compared to Model G. In other words, when cored halo profiles are assumed, the $f(R)$ signal at $\log_{10}\afR \sim -7$ largely disappears, so that models with a cored halo profile and a fifth force are not significantly better than those with a cored halo profile and no fifth force. 

All of this is not to say, however, that no galaxy prefers an $f(R)$ fifth force to a cored halo. We repeat the analysis shown in Figure \ref{F:dBIC_CDF}, i.e. the distribution of $\Delta\mathrm{BIC}$ for a freely varying $\fR$ model versus $\Lambda$CDM, but replacing the $\Lambda$CDM+NFW model with the $\Lambda$CDM+DC14 model (i.e. $\Delta\mathrm{BIC}\equiv \mathrm{BIC}_B - \mathrm{BIC}_G$). In this case, we find that it is now only a minority of galaxies that show \textit{some} preference for $f(R)$, with $29/85$ galaxies having $\Delta\mathrm{BIC}<0$, down from 64 in the NFW case. The numbers at the extremes have also shifted, with 14 galaxies showing a strongly significant ($|\Delta\mathrm{BIC}|>6$) preference for $f(R)$ and 21 for $\Lambda$CDM, compared with 27 and 0 respectively in the NFW case. These numbers reinforce the idea conveyed by Figure \ref{F:lnLRatios}: $\Lambda$CDM with a cored halo profile appears to be preferable over a cuspy halo and $f(R)$ for most of the galaxies in the sample.

Figure \ref{F:BvG} shows rotation curve fits of the 3 galaxies with the most extreme preferences for $f(R)$ over $\Lambda$CDM+DC14 (i.e. the three galaxies with the most negative values of $\Delta\mathrm{BIC}\equiv \mathrm{BIC}_B - \mathrm{BIC}_G$). These are UGC11820 ($\Delta\mathrm{BIC} = -22.1$), NGC2403 (-31.2), and UGC05253 (-47.1). In the case of UGC11820, it appears that the $f(R)$ model does indeed do a markedly better job of fitting the rotation curve. There is not a visible `upturn' feature in the rotation curve, but the the overall shape of the rotation curve is much better described by an NFW halo enchanced with a fifth force in the outer regions than by a cored dark matter halo. Indeed, considering the goodness of fit in each case, the reduced chi-squared statistic $\chi^2_\nu$, drops from 13.4 to 0.8. 

However, in the case of UGC05253, it would appear that there are large-scale features in the rotation curve, both in the central (magnified) region and in the outer stretches, that are not adequately captured by our baryonic models. This picture is borne out by the values of $\chi^2_\nu$: 17.7 for Model G, and 9.5 for Model B. The screening radius is being imposed to markedly improve the quality of the fits, but in actuality neither model does particularly well. NGC2403 shows similar behaviour, although perhaps not quite so starkly. A significant upturn feature at $\sim 4$ kpc is clearly visible. Both models, with and without a screening radius, approximate this feature. However, the baryonic models are unable to fully fit its magnitude, so a screening radius is imposed to bridge the gap. Again, neither model performs particularly well: $\chi^2_\nu$ drops from 5.5 to 2.6.

In addition to these caveats, it should be once again borne in mind that the galaxies preferring Model B to Model G are not necessarily showing any preference for a single global $\fR$ value. In fact, the inferred values of $\log_{10}\afR$ for the extreme galaxies shown in Figure \ref{F:BvG} are rather discrepant: -7.07 for UGC11820, -6.57 for NGC2403, and -6.19 for UGC05253. As shown in Figure \ref{F:lnLRatios}, no single, global $\fR$ model with a NFW profile is in better agreement with the data than a $\Lambda$CDM model with a cored DC14 profile.

\subsection{Environmental Screening}
\label{S:Results:EnvScreening}

As described in \S\ref{S:Methods:EnvScreening}, we investigate a model with freely varying $\fR$, NFW haloes, and an additional large-scale overdensity to account for the environmental contribution to the scalar field profile. This is Model C, which is identical to Model B in every respect other than the presence of this large-scale overdensity. For a given $\fR$ as drawn in the Monte Carlo sampling, the structure will have gravitational potential equal to $\Phi_\mathrm{ext}$ calculated from the screening maps of \citet{Desmond2018c}, and size equal to the Compton wavelength calculated using Eq. (\ref{E:Compton}).

\begin{figure}
    \centering
    \includegraphics{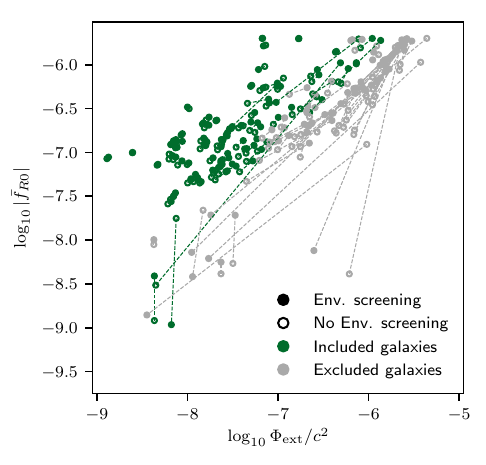}
    \caption{Best fit $\fR$ values for an $f(R)$ model excluding an external large-scale structure (i.e. Model B; unfilled circles) and including one (Model C; filled circles). The two points for a given galaxy are joined by a dashed line to indicate the change caused by including this environmental contribution. Green circles represent the 85 galaxies from our final sample, while grey circles are the 62 galaxies that were cut from the sample due to their overpopulated environment (see text for further details). The $x$-value for a given point indicates the $\Phi_\mathrm{ext}$ calculated from the screening map of \citet{Desmond2018c} for that given galaxy at the corresponding $\fR$ value. This figure indicates that the effect of environmental screening is primarily significant for galaxies that have been cut from the sample, while the environmental effects on our results are small in the galaxies that constitute our final sample.}
    \label{F:EnvScreening}
\end{figure}

Figure \ref{F:EnvScreening} shows a comparison of the $\fR$ values inferred from the maximum of the posterior of models Models B and C. Results for the 85 galaxies in our sample are shown, as well as results for the 62 galaxies that were excluded from the sample because of their significant environmental contribution.

Upon including the environmental contribution, the majority of the galaxies in Figure \ref{F:EnvScreening} move towards higher $\afR$ and to higher $\Phi_\mathrm{ext}$. This makes physical sense: for a given $f_{R0}$, the addition of an overdense environment will move the screening radius of a galaxy outward. For a rotation curve fit that is improved by a screening radius at a specific location, this effect is compensated by a higher value of $\afR$, hence the shift towards higher $\afR$. The shift towards higher $\Phi_\mathrm{ext}$ is due to the implicit relation between $\fR$ and $\Phi_\mathrm{ext}$: $\Phi_\mathrm{ext}$ is integrated up to the Compton wavelength, which in turn is determined by Eq. (\ref{E:Compton}). A higher $\afR$ will therefore lead to a higher $\Phi_\mathrm{ext}$ (note that this is also the reason for the correlation between inferred $\fR$ and $\Phi_\mathrm{ext}$ evident in Figure \ref{F:EnvScreening}). For these cases, this change is visually much larger for the excluded (environmentally screened) galaxies than for the included galaxies.

There is however, also a subset of galaxies for which the jumps are rather large, and these appear to be equally likely to be towards higher or lower $\afR$. In these cases, $f_{R0}$ is typically very poorly constrained and the marginal posteriors are rather broad, flat distributions (both with and without the environmental contribution). In such scenarios, the peak of the posterior can undergo quite large shifts (in either direction) between models, but the distributions are nonetheless mutually consistent.

The finding that the excluded galaxies exhibit a large change in inferred $\fR$ (median $|\Delta\log_{10}\afR|=0.37$) gives a post hoc justification for their exclusion from the sample. Meanwhile, the finding that the included galaxies typically do not change their inferred $\fR$ values significantly (median $|\Delta\log_{10}\afR|=0.04$) gives a post hoc justification for our neglecting the environmental screening effect in most of our models (see Table \ref{T:Models}).

\subsection{Stellar Self-Screening}
\label{S:Results:StellarScreening}

\begin{figure}
    \centering
    \includegraphics{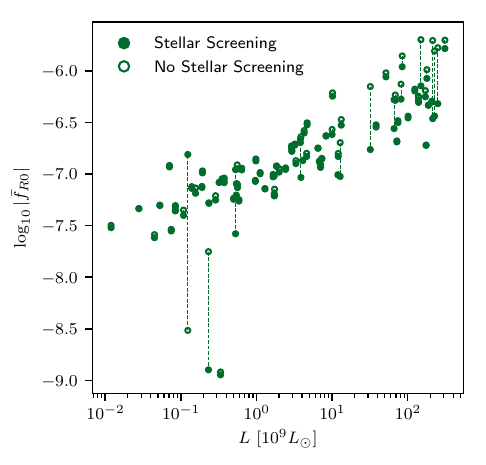}
    \caption{Best fit $\fR$ values for all 85 galaxies, from a model with stellar self-screening (Model D; filled circles) and without (Model B; unfilled circles), as a function of total stellar luminosity. As with Figure \ref{F:EnvScreening}, the small changes shown in this figure indicate that the effect of stellar self-screening on our results is small for the vast majority of the galaxies in our final sample. See the text for a discussion of the few large outliers.}
    \label{F:StellarScreening}
\end{figure}

Figure \ref{F:StellarScreening} shows a comparison of the $\fR$ values from the best fit models for all galaxies under Models D and B, i.e. $f(R)$ models with and without stellar self-screening, as a function of total stellar luminosity. In Model D, stars do not act as a source of the fifth force, as would be expected if they self-screen completely. Self-screening would also prevent a fifth force from accelerating stars, this is however not relevant here as the measured rotation curves are based on observations of the gas component.

The mean (absolute) change in $|\Delta\log_{10}\afR|$ between Models D and B is 0.11, but this number is dominated by a few outliers. In general, the change is not too large (median $|\Delta\log_{10}\afR|=0.02$), with a larger change taking place for more luminous galaxies.

The galaxies with the two largest changes in inferred $\fR$ are UGC05918 ($|\Delta\log_{10}\afR|=1.1$) and UGC07866 (1.7). In the case of UGC05918, under both Models B and D, the best fit models are those in which the galaxy is fully screened. In such a scenario, $\fR$ is poorly constrained as any $|\fR|$ value below a certain threshold will fully screen the galaxy, and all such values have equal posterior probability.

Meanwhile, in the case of UGC07866, the posterior in both Models B and D is bimodal. In one mode, the galaxy is fully screened, while in the other mode, the galaxy is fully unscreened, and the mass of the halo and stellar components are renormalised by a factor of 3/4 to give a very similar rotation curve model. Assuming stellar self-screening causes a slight shift in the relative weights of the two modes, which caused the best fit model to spontaneously `jump' from one mode to the other.

Neither of these cases, nor indeed the other outliers with large changes in inferred $\fR$, should be causes for concern. In each of these cases, the best fit model either fully screens or fully unscreens the galaxy. So, while the best fit values are different under the two models, the marginal posterior distributions for $\fR$ in both cases are wide, flat distributions that are consistent with each other. The majority of galaxies, however, exhibit the more typical behaviour, where a screening radius in a particular location is favourable, which corresponds to a specific value of $\fR$. For these galaxies, the required value of $\fR$ is very similar for Models B and D, which is reassuring and justifies neglecting the effect of stellar self-screening in most of the models listed in Table \ref{T:Models}.

\subsection{Mass-to-light Ratios}
\label{S:Results:MLRatios}

As described in \S\ref{S:Methods:Models}, Model B (as well as all other models considered besides E and F), performs fits with a single freely varying $\Upsilon$, which applies to both the bulge and disc components. Model E instead fixes $\Upsilon$ to be equal to  $0.5 M_\odot/L_\odot$ for the disc and 0.7 for the bulge, while Model F fits two freely varying ratios: $\Upsilon_\mathrm{disc}$ and $\Upsilon_\mathrm{bulge}$.

It should be noted that of the 85 galaxies in the sample, only 13 have a detected bulge component. For the remaining 72 galaxies, Models B and F are identical, and should therefore give identical results.

\begin{figure}
    \includegraphics{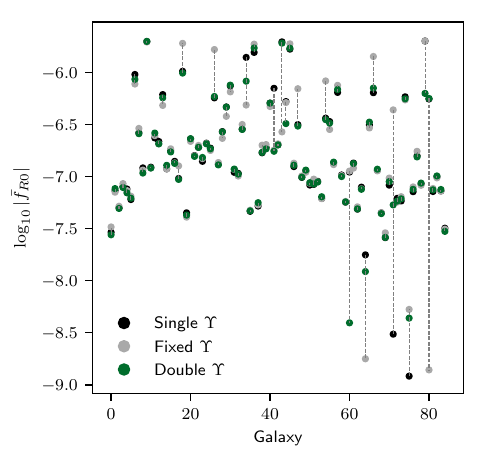}
    \caption{For each galaxy in the sample, $\fR$ values from the best-fit model, using three different treatments of the mass-to-light ratio $\Upsilon$ (i.e., Models B, E, and F). Purple points show the results from the model with one free parameter for $\Upsilon$ (Model B), while light and dark green points represent Models E and F: models with fixed empirical values, and with two free parameters (for bulge and disc) respectively. Grey dashed lines show the changes in the $\fR$ values between models for a given galaxy. This figure indicates that there is no significant change to the results in going from two free parameters to one, but fixing the values as empirical constants induces a more significant change.}
    \label{F:MLRatios}
\end{figure}

Figure \ref{F:MLRatios} shows a comparison of results from Models B, E, and F. For all 85 galaxies in the sample, the figure shows the $\fR$ value for the best fit model, for the B, E and F models. The mean (absolute) change is $|\Delta\log_{10}\afR|=0.14$ between Models B and E, but this is dominated by a few outliers, and the median change is 0.03. For Models B and F, considering only the 13 galaxies with bulge components, the mean change is 0.07 and the median is 0.01.

These results suggest that fixing the values of the mass-to-light ratios might be too crude an approximation. Indeed, most of the visibly large changes in Figure \ref{F:MLRatios} are due to Model E. On the other hand, the reassuringly small changes between Models B and F imply that taking the stellar disc and bulge to have the same mass-to-light ratio is a sufficient approximation.

% CONCLUSION
\section{Discussion and Conclusions}
\label{S:Conclusions}

In this work, we have searched for potential signatures of Chameleon-$f(R)$ modified gravity in the high-quality HI/H$\alpha$ rotation curve measurements of the SPARC sample. After several cuts to the sample, including an exclusion of all galaxies likely to be environmentally screened, 85 of the original 175 galaxies remained.

Each galaxy is modelled as a stellar disc, gaseous disc, dark matter halo, and (where appropriate) a stellar bulge. Then, for a given $\fR$ value, we solve the $f(R)$ equations assuming spherical symmetry, and calculate the implied fifth force contribution to the rotation curve. Assuming a Gaussian likelihood, we are then able to use an MCMC technique to explore suitable models and search for evidence of modified gravity.

As mentioned in the Introduction, all of the code, analysis tools, and plotting scripts used in this work have been made into a publicly available \texttt{python 3} package\footnote{\href{https://github.com/aneeshnaik/spam}{https://github.com/aneeshnaik/spam}}.

Our main findings are:
\begin{itemize}
    \item In models with an NFW halo and an $\fR$ that is allowed to vary freely for each galaxy, most of the galaxy sample (64/85) indicated \textit{some} preference for $f(R)$, according to the Bayesian Information Criterion. Furthermore, 27 galaxies showed a strongly significant ($|\Delta\mathrm{BIC}|>6$) preference for $f(R)$.
    \item Looking at the marginal posterior distributions of $\fR$ of all galaxies in our final sample, most galaxies have reasonably tight posteriors and $\fR$ appears well constrained, but the spread of these inferred $\fR$ values across the sample is very broad, spanning roughly two orders of magnitude. This is inconsistent with a single global value for $\fR$.
    \item This finding is confirmed when analysing models with fixed, globally imposed $\fR$ values. Models with $\log_{10}\afR \gtrsim -6.1$ are highly unfavourable compared to $\Lambda$CDM. However, models with lower $\afR$ appear to be preferred over $\Lambda$CDM. This signal reaches a peak at $\log_{10}\afR \sim -7$, where the log-likelihood ratio $\ln\mathcal{L}_{f(R)}/\mathcal{L}_{\Lambda\mathrm{CDM}} \sim 70$.
    \item The galaxies that dominate this signal are dwarf galaxies, which have previously been noted to exhibit the core-cusp problem.
    \item Modelling with cored DC14 haloes without a fifth force, the overall log-likelihood shows an increase of nearly 300, far more significant than any $f(R)$-model with a single, global $\fR$ value.
    \item Modelling with cored DC14 haloes with a fifth force does not provide a significant improvement over the case of DC14+$\Lambda$CDM, and the signal at $\afR \sim 10^{-7}$ has largely vanished.
    \item There are nonetheless some individual galaxies which show a preference for NFW+$f(R)$ over DC14+$\Lambda$CDM. However, for some of these galaxies it appears to be the case that neither model performs particularly well, as a result of large-scale baryonic features (e.g., due to prominent spiral arms) in the rotation curves inadequately captured by our baryonic models.
    \item These results appear to be robust to the effects of environmental screening. Having removed a significant fraction of the galaxies for which environmental screening was believed to play a significant role, the results of the $\fR$ inference for the remaining 85 galaxies were not affected significantly when a large-scale environment was added to the modelling pipeline.
    \item Similar analysis was performed to test for the effects of stellar self-screening and a more general treatment of the mass to light ratio. It was also found here that the results of our inference are robust to these effects.
\end{itemize}

We thus end by reporting an absence of convincing evidence of modified gravity down to the sensitivity level of our sample at $\afR \sim 6 \times 10^{-8}$, as the improvements to the rotation curve fits due to the $f(R)$ fifth force (which peak at around $\afR \sim 10^{-7}$) are more readily explained by galaxies having cored dark matter profiles than by a genuine signal of modified gravity.

The only work, to our knowledge, that has previously addressed this degeneracy between cored dark matter profiles and screened modified gravity theories is that of \citet{Lombriser2015}, which considered the velocity dispersions of local dwarf spheroidal galaxies Fornax and Sculptor. The radial slopes of their velocity dispersions have previously been shown to be consistent with cored mass profiles \citep{Walker2011}, and \citet{Lombriser2015} show that they can alternatively be interpreted as being due to a chameleon fifth force which is unscreened only in the outer parts of these dwarfs. However, the $\fR = -10^{-7}$ models that we prefer may not be suitable for this as they would likely, depending on the exact Milky Way mass and the environmental effects of the Local Group, screen the Milky Way halo to beyond the positions of the Fornax and Sculptor dwarfs.

It would therefore be interesting to extend the remit of the present work by investigating a wider range of screened modified gravity theories to investigate the constraints that can be derived on them, and whether any gravity model can bring cuspy NFW profiles into good agreement with galaxy rotation curve data.

% ACKNOWLEDGEMENTS
\section*{Acknowledgements}

We would like to thank Federico Lelli, Harley Katz, Cameron Lemon, and the anonymous reviewer for helpful comments and discussions.

APN thanks the Science and Technology Facilities Council (STFC) for their PhD studentship. EP acknowledges support by the Kavli foundation. ACD acknowledges partial support from STFC under grants ST/L000385 and ST/L000636. DS acknowledges ERC starting grant 638707 and support from the STFC. HD is supported by St John’s College, Oxford and acknowledges financial support from ERC Grant No 693024 and the Beecroft Trust.

This work used the DiRAC (www.dirac.ac.uk) system: Data Analytic at the University of Cambridge [funded by BIS National E-infrastructure capital grant (ST/K001590/1), STFC capital grants ST/H008861/1 and ST/H00887X/1, and STFC DiRAC Operations grant ST/K00333X/1]. DiRAC is part of the National E-Infrastructure.

% REFERENCES
\bibliographystyle{mnras}
\bibliography{library}

\begin{thebibliography}{}
\makeatletter
\relax
\def\mn@urlcharsother{\let\do\@makeother \do\$\do\&\do\#\do\^\do\_\do\%\do\~}
\def\mn@doi{\begingroup\mn@urlcharsother \@ifnextchar [ {\mn@doi@}
  {\mn@doi@[]}}
\def\mn@doi@[#1]#2{\def\@tempa{#1}\ifx\@tempa\@empty \href
  {http://dx.doi.org/#2} {doi:#2}\else \href {http://dx.doi.org/#2} {#1}\fi
  \endgroup}
\def\mn@eprint#1#2{\mn@eprint@#1:#2::\@nil}
\def\mn@eprint@arXiv#1{\href {http://arxiv.org/abs/#1} {{\tt arXiv:#1}}}
\def\mn@eprint@dblp#1{\href {http://dblp.uni-trier.de/rec/bibtex/#1.xml}
  {dblp:#1}}
\def\mn@eprint@#1:#2:#3:#4\@nil{\def\@tempa {#1}\def\@tempb {#2}\def\@tempc
  {#3}\ifx \@tempc \@empty \let \@tempc \@tempb \let \@tempb \@tempa \fi \ifx
  \@tempb \@empty \def\@tempb {arXiv}\fi \@ifundefined
  {mn@eprint@\@tempb}{\@tempb:\@tempc}{\expandafter \expandafter \csname
  mn@eprint@\@tempb\endcsname \expandafter{\@tempc}}}

\bibitem[\protect\citeauthoryear{{Amendola} \& {Tsujikawa}}{{Amendola} \&
  {Tsujikawa}}{2010}]{Amendola2010}
{Amendola} L.,  {Tsujikawa} S.,  2010, {Dark Energy: Theory and Observations}.
Cambridge University Press

\bibitem[\protect\citeauthoryear{{Arnold}, {Fosalba}, {Springel}, {Puchwein}
  \& {Blot}}{{Arnold} et~al.}{2019}]{Arnold2019}
{Arnold} C.,  {Fosalba} P.,  {Springel} V.,  {Puchwein} E.,   {Blot} L.,  2019,
  \mn@doi [\mnras] {10.1093/mnras/sty3044}, \href
  {https://ui.adsabs.harvard.edu/abs/2019MNRAS.483..790A} {483, 790}

\bibitem[\protect\citeauthoryear{{Bose} et~al.,}{{Bose}
  et~al.}{2019}]{Bose2018}
{Bose} S.,  et~al., 2019, \mn@doi [\mnras] {10.1093/mnras/stz1168}, \href
  {https://ui.adsabs.harvard.edu/abs/2019MNRAS.486.4790B} {486, 4790}

\bibitem[\protect\citeauthoryear{{Boylan-Kolchin}, {Bullock}  \&
  {Kaplinghat}}{{Boylan-Kolchin} et~al.}{2011}]{BoylanKolchin2011}
{Boylan-Kolchin} M.,  {Bullock} J.~S.,   {Kaplinghat} M.,  2011, \mn@doi
  [\mnras] {10.1111/j.1745-3933.2011.01074.x}, \href
  {http://ukads.nottingham.ac.uk/abs/2011MNRAS.415L..40B} {415, L40}

\bibitem[\protect\citeauthoryear{{Brax}, {van de Bruck}, {Davis}  \&
  {Shaw}}{{Brax} et~al.}{2008}]{Brax2008}
{Brax} P.,  {van de Bruck} C.,  {Davis} A.-C.,   {Shaw} D.~J.,  2008, \mn@doi
  [\prd] {10.1103/PhysRevD.78.104021}, \href
  {http://ukads.nottingham.ac.uk/abs/2008PhRvD..78j4021B} {78, 104021}

\bibitem[\protect\citeauthoryear{{Buchdahl}}{{Buchdahl}}{1970}]{Buchdahl1970}
{Buchdahl} H.~A.,  1970, \mn@doi [\mnras] {10.1093/mnras/150.1.1}, \href
  {http://ukads.nottingham.ac.uk/abs/1970MNRAS.150....1B} {150, 1}

\bibitem[\protect\citeauthoryear{{Bull} et~al.,}{{Bull}
  et~al.}{2016}]{Bull2016}
{Bull} P.,  et~al., 2016, \mn@doi [Physics of the Dark Universe]
  {10.1016/j.dark.2016.02.001}, \href
  {https://ui.adsabs.harvard.edu/\#abs/2016PDU....12...56B} {12, 56}

\bibitem[\protect\citeauthoryear{{Bullock} \& {Boylan-Kolchin}}{{Bullock} \&
  {Boylan-Kolchin}}{2017}]{Bullock2017}
{Bullock} J.~S.,  {Boylan-Kolchin} M.,  2017, \mn@doi [Annual Review of
  Astronomy and Astrophysics] {10.1146/annurev-astro-091916-055313}, \href
  {https://ui.adsabs.harvard.edu/\#abs/2017ARA&A..55..343B} {55, 343}

\bibitem[\protect\citeauthoryear{{Burrage} \& {Sakstein}}{{Burrage} \&
  {Sakstein}}{2016}]{Burrage2016}
{Burrage} C.,  {Sakstein} J.,  2016, \mn@doi [Journal of Cosmology and
  Astro-Particle Physics] {10.1088/1475-7516/2016/11/045}, \href
  {https://ui.adsabs.harvard.edu/\#abs/2016JCAP...11..045B} {2016, 045}

\bibitem[\protect\citeauthoryear{{Burrage} \& {Sakstein}}{{Burrage} \&
  {Sakstein}}{2018}]{Burrage2018}
{Burrage} C.,  {Sakstein} J.,  2018, \mn@doi [Living Reviews in Relativity]
  {10.1007/s41114-018-0011-x}, \href
  {http://ukads.nottingham.ac.uk/abs/2018LRR....21....1B} {21, 1}

\bibitem[\protect\citeauthoryear{{Cabr{\'e}}, {Vikram}, {Zhao}, {Jain}  \&
  {Koyama}}{{Cabr{\'e}} et~al.}{2012}]{Cabre2012}
{Cabr{\'e}} A.,  {Vikram} V.,  {Zhao} G.-B.,  {Jain} B.,   {Koyama} K.,  2012,
  \mn@doi [\jcap] {10.1088/1475-7516/2012/07/034}, \href
  {http://ukads.nottingham.ac.uk/abs/2012JCAP...07..034C} {7, 034}

\bibitem[\protect\citeauthoryear{{Cataneo}, {Rapetti}, {Lombriser}  \&
  {Li}}{{Cataneo} et~al.}{2016}]{Cataneo2016}
{Cataneo} M.,  {Rapetti} D.,  {Lombriser} L.,   {Li} B.,  2016, \mn@doi
  [Journal of Cosmology and Astro-Particle Physics]
  {10.1088/1475-7516/2016/12/024}, \href
  {https://ui.adsabs.harvard.edu/\#abs/2016JCAP...12..024C} {2016, 024}

\bibitem[\protect\citeauthoryear{{Clifton}, {Ferreira}, {Padilla}  \&
  {Skordis}}{{Clifton} et~al.}{2012}]{Clifton2012}
{Clifton} T.,  {Ferreira} P.~G.,  {Padilla} A.,   {Skordis} C.,  2012, \mn@doi
  [\physrep] {10.1016/j.physrep.2012.01.001}, \href
  {http://adsabs.harvard.edu/abs/2012PhR...513....1C} {513, 1}

\bibitem[\protect\citeauthoryear{{Davis}, {Lim}, {Sakstein}  \& {Shaw}}{{Davis}
  et~al.}{2012}]{Davis2012}
{Davis} A.-C.,  {Lim} E.~A.,  {Sakstein} J.,   {Shaw} D.~J.,  2012, \mn@doi
  [\prd] {10.1103/PhysRevD.85.123006}, \href
  {http://adsabs.harvard.edu/abs/2012PhRvD..85l3006D} {85, 123006}

\bibitem[\protect\citeauthoryear{{Desmond}}{{Desmond}}{2017a}]{Desmond2017a}
{Desmond} H.,  2017a, \mn@doi [\mnras] {10.1093/mnras/stw2571}, \href
  {https://ui.adsabs.harvard.edu/abs/2017MNRAS.464.4160D} {464, 4160}

\bibitem[\protect\citeauthoryear{{Desmond}}{{Desmond}}{2017b}]{Desmond2017b}
{Desmond} H.,  2017b, \mn@doi [\mnras] {10.1093/mnrasl/slx134}, \href
  {https://ui.adsabs.harvard.edu/abs/2017MNRAS.472L..35D} {472, L35}

\bibitem[\protect\citeauthoryear{{Desmond}, {Ferreira}, {Lavaux}  \&
  {Jasche}}{{Desmond} et~al.}{2018a}]{Desmond2018a}
{Desmond} H.,  {Ferreira} P.~G.,  {Lavaux} G.,   {Jasche} J.,  2018a, \mn@doi
  [\prd] {10.1103/PhysRevD.98.083010}, \href
  {https://ui.adsabs.harvard.edu/\#abs/2018PhRvD..98h3010D} {98, 083010}

\bibitem[\protect\citeauthoryear{{Desmond}, {Ferreira}, {Lavaux}  \&
  {Jasche}}{{Desmond} et~al.}{2018b}]{Desmond2018b}
{Desmond} H.,  {Ferreira} P.~G.,  {Lavaux} G.,   {Jasche} J.,  2018b, \mn@doi
  [\prd] {10.1103/PhysRevD.98.064015}, \href
  {https://ui.adsabs.harvard.edu/abs/2018PhRvD..98f4015D} {98, 064015}

\bibitem[\protect\citeauthoryear{{Desmond}, {Ferreira}, {Lavaux}  \&
  {Jasche}}{{Desmond} et~al.}{2018c}]{Desmond2018c}
{Desmond} H.,  {Ferreira} P.~G.,  {Lavaux} G.,   {Jasche} J.,  2018c, \mn@doi
  [\mnras] {10.1093/mnras/stx3062}, \href
  {https://ui.adsabs.harvard.edu/\#abs/2018MNRAS.474.3152D} {474, 3152}

\bibitem[\protect\citeauthoryear{{Desmond}, {Ferreira}, {Lavaux}  \&
  {Jasche}}{{Desmond} et~al.}{2019}]{Desmond2019}
{Desmond} H.,  {Ferreira} P.~G.,  {Lavaux} G.,   {Jasche} J.,  2019, \mn@doi
  [\mnras] {10.1093/mnrasl/sly221}, \href
  {https://ui.adsabs.harvard.edu/\#abs/2019MNRAS.483L..64D} {483, L64}

\bibitem[\protect\citeauthoryear{{Di Cintio}, {Brook}, {Dutton}, {Macci{\`o}},
  {Stinson}  \& {Knebe}}{{Di Cintio} et~al.}{2014}]{DiCintio2014}
{Di Cintio} A.,  {Brook} C.~B.,  {Dutton} A.~A.,  {Macci{\`o}} A.~V.,
  {Stinson} G.~S.,   {Knebe} A.,  2014, \mn@doi [\mnras]
  {10.1093/mnras/stu729}, \href
  {https://ui.adsabs.harvard.edu/\#abs/2014MNRAS.441.2986D} {441, 2986}

\bibitem[\protect\citeauthoryear{{Dutton} \& {Macci{\`o}}}{{Dutton} \&
  {Macci{\`o}}}{2014}]{Dutton2014}
{Dutton} A.~A.,  {Macci{\`o}} A.~V.,  2014, \mn@doi [\mnras]
  {10.1093/mnras/stu742}, \href
  {http://ukads.nottingham.ac.uk/abs/2014MNRAS.441.3359D} {441, 3359}

\bibitem[\protect\citeauthoryear{{Ezquiaga} \& {Zumalac{\'a}rregui}}{{Ezquiaga}
  \& {Zumalac{\'a}rregui}}{2017}]{Maria2017}
{Ezquiaga} J.~M.,  {Zumalac{\'a}rregui} M.,  2017, \mn@doi [\prl]
  {10.1103/PhysRevLett.119.251304}, \href
  {https://ui.adsabs.harvard.edu/abs/2017PhRvL.119y1304E} {119, 251304}

\bibitem[\protect\citeauthoryear{{Flores} \& {Primack}}{{Flores} \&
  {Primack}}{1994}]{Flores1994}
{Flores} R.~A.,  {Primack} J.~R.,  1994, \mn@doi [\apj] {10.1086/187350}, \href
  {https://ui.adsabs.harvard.edu/\#abs/1994ApJ...427L...1F} {427, L1}

\bibitem[\protect\citeauthoryear{{Foreman-Mackey}, {Hogg}, {Lang}  \&
  {Goodman}}{{Foreman-Mackey} et~al.}{2013}]{Foreman2013}
{Foreman-Mackey} D.,  {Hogg} D.~W.,  {Lang} D.,   {Goodman} J.,  2013, \mn@doi
  [\pasp] {10.1086/670067}, \href
  {http://adsabs.harvard.edu/abs/2013PASP..125..306F} {125, 306}

\bibitem[\protect\citeauthoryear{{Gelman} \& {Rubin}}{{Gelman} \&
  {Rubin}}{1992}]{Gelman1992}
{Gelman} A.,  {Rubin} D.~B.,  1992, \mn@doi [Statistical Science]
  {10.1214/ss/1177011136}, \href
  {http://ukads.nottingham.ac.uk/abs/1992StaSc...7..457G} {7, 457}

\bibitem[\protect\citeauthoryear{{Genina} et~al.,}{{Genina}
  et~al.}{2018}]{Genina2018}
{Genina} A.,  et~al., 2018, \mn@doi [\mnras] {10.1093/mnras/stx2855}, \href
  {https://ui.adsabs.harvard.edu/\#abs/2018MNRAS.474.1398G} {474, 1398}

\bibitem[\protect\citeauthoryear{{Governato} et~al.,}{{Governato}
  et~al.}{2010}]{Governato2010}
{Governato} F.,  et~al., 2010, \mn@doi [\nat] {10.1038/nature08640}, \href
  {http://adsabs.harvard.edu/abs/2010Natur.463..203G} {463, 203}

\bibitem[\protect\citeauthoryear{{Hu} \& {Sawicki}}{{Hu} \&
  {Sawicki}}{2007}]{Hu2007}
{Hu} W.,  {Sawicki} I.,  2007, \mn@doi [\prd] {10.1103/PhysRevD.76.064004},
  \href {http://ukads.nottingham.ac.uk/abs/2007PhRvD..76f4004H} {76, 064004}

\bibitem[\protect\citeauthoryear{{Jain} \& {Khoury}}{{Jain} \&
  {Khoury}}{2010}]{Jain2010}
{Jain} B.,  {Khoury} J.,  2010, \mn@doi [Annals of Physics]
  {10.1016/j.aop.2010.04.002}, \href
  {http://ukads.nottingham.ac.uk/abs/2010AnPhy.325.1479J} {325, 1479}

\bibitem[\protect\citeauthoryear{{Jain}, {Vikram}  \& {Sakstein}}{{Jain}
  et~al.}{2013}]{Jain2013}
{Jain} B.,  {Vikram} V.,   {Sakstein} J.,  2013, \mn@doi [\apj]
  {10.1088/0004-637X/779/1/39}, \href
  {http://ukads.nottingham.ac.uk/abs/2013ApJ...779...39J} {779, 39}

\bibitem[\protect\citeauthoryear{{Joyce}, {Jain}, {Khoury}  \&
  {Trodden}}{{Joyce} et~al.}{2015}]{Joyce2015}
{Joyce} A.,  {Jain} B.,  {Khoury} J.,   {Trodden} M.,  2015, \mn@doi [\physrep]
  {10.1016/j.physrep.2014.12.002}, \href
  {http://ukads.nottingham.ac.uk/abs/2015PhR...568....1J} {568, 1}

\bibitem[\protect\citeauthoryear{{Kass} \& {Raftery}}{{Kass} \&
  {Raftery}}{1995}]{Kass1995}
{Kass} R.~E.,  {Raftery} A.~E.,  1995, \mn@doi [Journal of the American
  Statistical Association] {10.1080/01621459.1995.10476572}, 90, 773

\bibitem[\protect\citeauthoryear{{Katz}, {Lelli}, {McGaugh}, {Di Cintio},
  {Brook}  \& {Schombert}}{{Katz} et~al.}{2017}]{Katz2017}
{Katz} H.,  {Lelli} F.,  {McGaugh} S.~S.,  {Di Cintio} A.,  {Brook} C.~B.,
  {Schombert} J.~M.,  2017, \mn@doi [\mnras] {10.1093/mnras/stw3101}, \href
  {http://adsabs.harvard.edu/abs/2017MNRAS.466.1648K} {466, 1648}

\bibitem[\protect\citeauthoryear{{Katz}, {Desmond}, {Lelli}, {McGaugh}, {Di
  Cintio}, {Brook}  \& {Schombert}}{{Katz} et~al.}{2018}]{Katz2018}
{Katz} H.,  {Desmond} H.,  {Lelli} F.,  {McGaugh} S.,  {Di Cintio} A.,  {Brook}
  C.,   {Schombert} J.,  2018, \mn@doi [\mnras] {10.1093/mnras/sty2129}, \href
  {https://ui.adsabs.harvard.edu/\#abs/2018MNRAS.480.4287K} {480, 4287}

\bibitem[\protect\citeauthoryear{{Khoury}}{{Khoury}}{2010}]{Khoury2010}
{Khoury} J.,  2010, arXiv e-prints, \href
  {http://ukads.nottingham.ac.uk/abs/2010arXiv1011.5909K} {p. arXiv:1011.5909}

\bibitem[\protect\citeauthoryear{{Khoury} \& {Weltman}}{{Khoury} \&
  {Weltman}}{2004}]{Khoury2004}
{Khoury} J.,  {Weltman} A.,  2004, \mn@doi [\prd] {10.1103/PhysRevD.69.044026},
  \href {http://ukads.nottingham.ac.uk/abs/2004PhRvD..69d4026K} {69, 044026}

\bibitem[\protect\citeauthoryear{{Klypin}, {Kravtsov}, {Valenzuela}  \&
  {Prada}}{{Klypin} et~al.}{1999}]{Klypin1999}
{Klypin} A.,  {Kravtsov} A.~V.,  {Valenzuela} O.,   {Prada} F.,  1999, \mn@doi
  [\apj] {10.1086/307643}, \href
  {https://ui.adsabs.harvard.edu/\#abs/1999ApJ...522...82K} {522, 82}

\bibitem[\protect\citeauthoryear{{Koyama}}{{Koyama}}{2016}]{Koyama2016}
{Koyama} K.,  2016, \mn@doi [Reports on Progress in Physics]
  {10.1088/0034-4885/79/4/046902}, \href
  {http://adsabs.harvard.edu/abs/2016RPPh...79d6902K} {79, 046902}

\bibitem[\protect\citeauthoryear{{Lelli}, {McGaugh}  \& {Schombert}}{{Lelli}
  et~al.}{2016}]{Lelli2016}
{Lelli} F.,  {McGaugh} S.~S.,   {Schombert} J.~M.,  2016, \mn@doi [\aj]
  {10.3847/0004-6256/152/6/157}, \href
  {http://adsabs.harvard.edu/abs/2016AJ....152..157L} {152, 157}

\bibitem[\protect\citeauthoryear{{Lelli}, {McGaugh}, {Schombert}  \&
  {Pawlowski}}{{Lelli} et~al.}{2017}]{Lelli2017}
{Lelli} F.,  {McGaugh} S.~S.,  {Schombert} J.~M.,   {Pawlowski} M.~S.,  2017,
  \mn@doi [\apj] {10.3847/1538-4357/836/2/152}, \href
  {http://adsabs.harvard.edu/abs/2017ApJ...836..152L} {836, 152}

\bibitem[\protect\citeauthoryear{{Li}, {Lelli}, {McGaugh}  \& {Schombert}}{{Li}
  et~al.}{2018}]{Li2018}
{Li} P.,  {Lelli} F.,  {McGaugh} S.,   {Schombert} J.,  2018, \mn@doi [\aap]
  {10.1051/0004-6361/201732547}, \href
  {http://adsabs.harvard.edu/abs/2018A%26A...615A...3L} {615, A3}

\bibitem[\protect\citeauthoryear{{Lombriser}}{{Lombriser}}{2014}]{Lombriser2014}
{Lombriser} L.,  2014, \mn@doi [Annalen der Physik] {10.1002/andp.201400058},
  \href {https://ui.adsabs.harvard.edu/\#abs/2014AnP...526..259L} {264, 259}

\bibitem[\protect\citeauthoryear{{Lombriser} \& {Lima}}{{Lombriser} \&
  {Lima}}{2017}]{Lombriser2017}
{Lombriser} L.,  {Lima} N.~A.,  2017, \mn@doi [Physics Letters B]
  {10.1016/j.physletb.2016.12.048}, \href
  {http://adsabs.harvard.edu/abs/2017PhLB..765..382L} {765, 382}

\bibitem[\protect\citeauthoryear{{Lombriser} \& {Pe{\~n}arrubia}}{{Lombriser}
  \& {Pe{\~n}arrubia}}{2015}]{Lombriser2015}
{Lombriser} L.,  {Pe{\~n}arrubia} J.,  2015, \mn@doi [\prd]
  {10.1103/PhysRevD.91.084022}, \href
  {https://ui.adsabs.harvard.edu/\#abs/2015PhRvD..91h4022L} {91, 084022}

\bibitem[\protect\citeauthoryear{{Mashchenko}, {Wadsley}  \&
  {Couchman}}{{Mashchenko} et~al.}{2008}]{Mashchenko2008}
{Mashchenko} S.,  {Wadsley} J.,   {Couchman} H.~M.~P.,  2008, \mn@doi [Science]
  {10.1126/science.1148666}, \href
  {https://ui.adsabs.harvard.edu/\#abs/2008Sci...319..174M} {319, 174}

\bibitem[\protect\citeauthoryear{{McGaugh} \& {Schombert}}{{McGaugh} \&
  {Schombert}}{2014}]{McGaugh2014}
{McGaugh} S.~S.,  {Schombert} J.~M.,  2014, \mn@doi [\aj]
  {10.1088/0004-6256/148/5/77}, \href
  {http://ukads.nottingham.ac.uk/abs/2014AJ....148...77M} {148, 77}

\bibitem[\protect\citeauthoryear{{McGaugh}, {Lelli}  \& {Schombert}}{{McGaugh}
  et~al.}{2016}]{McGaugh2016}
{McGaugh} S.~S.,  {Lelli} F.,   {Schombert} J.~M.,  2016, \mn@doi [Physical
  Review Letters] {10.1103/PhysRevLett.117.201101}, \href
  {http://adsabs.harvard.edu/abs/2016PhRvL.117t1101M} {117, 201101}

\bibitem[\protect\citeauthoryear{{Meidt} et~al.,}{{Meidt}
  et~al.}{2014}]{Meidt2014}
{Meidt} S.~E.,  et~al., 2014, \mn@doi [\apj] {10.1088/0004-637X/788/2/144},
  \href {http://ukads.nottingham.ac.uk/abs/2014ApJ...788..144M} {788, 144}

\bibitem[\protect\citeauthoryear{{Mitchell}, {Arnold}, {He}  \&
  {Li}}{{Mitchell} et~al.}{2019}]{Mitchell2019}
{Mitchell} M.~A.,  {Arnold} C.,  {He} J.-h.,   {Li} B.,  2019, \mn@doi [\mnras]
  {10.1093/mnras/stz1389}, \href
  {https://ui.adsabs.harvard.edu/abs/2019MNRAS.tmp.1334M} {p.~1334}

\bibitem[\protect\citeauthoryear{{Moore}}{{Moore}}{1994}]{Moore1994}
{Moore} B.,  1994, \mn@doi [\nat] {10.1038/370629a0}, \href
  {https://ui.adsabs.harvard.edu/\#abs/1994Natur.370..629M} {370, 629}

\bibitem[\protect\citeauthoryear{{Moore}, {Ghigna}, {Governato}, {Lake},
  {Quinn}, {Stadel}  \& {Tozzi}}{{Moore} et~al.}{1999}]{Moore1999}
{Moore} B.,  {Ghigna} S.,  {Governato} F.,  {Lake} G.,  {Quinn} T.,  {Stadel}
  J.,   {Tozzi} P.,  1999, \mn@doi [\apj] {10.1086/312287}, \href
  {https://ui.adsabs.harvard.edu/\#abs/1999ApJ...524L..19M} {524, L19}

\bibitem[\protect\citeauthoryear{{Moster}, {Naab}  \& {White}}{{Moster}
  et~al.}{2013}]{Moster2013}
{Moster} B.~P.,  {Naab} T.,   {White} S.~D.~M.,  2013, \mn@doi [\mnras]
  {10.1093/mnras/sts261}, \href
  {http://ukads.nottingham.ac.uk/abs/2013MNRAS.428.3121M} {428, 3121}

\bibitem[\protect\citeauthoryear{{Naik}, {Puchwein}, {Davis}  \&
  {Arnold}}{{Naik} et~al.}{2018}]{Naik2018}
{Naik} A.~P.,  {Puchwein} E.,  {Davis} A.-C.,   {Arnold} C.,  2018, \mn@doi
  [\mnras] {10.1093/mnras/sty2199}, \href
  {http://ukads.nottingham.ac.uk/abs/2018MNRAS.480.5211N} {480, 5211}

\bibitem[\protect\citeauthoryear{{Navarro}, {Frenk}  \& {White}}{{Navarro}
  et~al.}{1997}]{Navarro1997}
{Navarro} J.~F.,  {Frenk} C.~S.,   {White} S. D.~M.,  1997, \mn@doi [\apj]
  {10.1086/304888}, \href
  {https://ui.adsabs.harvard.edu/\#abs/1997ApJ...490..493N} {490, 493}

\bibitem[\protect\citeauthoryear{{Noller}, {von Braun-Bates}  \&
  {Ferreira}}{{Noller} et~al.}{2014}]{Noller2014}
{Noller} J.,  {von Braun-Bates} F.,   {Ferreira} P.~G.,  2014, \mn@doi [\prd]
  {10.1103/PhysRevD.89.023521}, \href
  {http://ukads.nottingham.ac.uk/abs/2014PhRvD..89b3521N} {89, 023521}

\bibitem[\protect\citeauthoryear{{Oh} et~al.,}{{Oh} et~al.}{2015}]{Oh2015}
{Oh} S.-H.,  et~al., 2015, \mn@doi [\aj] {10.1088/0004-6256/149/6/180}, \href
  {https://ui.adsabs.harvard.edu/\#abs/2015AJ....149..180O} {149, 180}

\bibitem[\protect\citeauthoryear{{Planck Collaboration}}{{Planck
  Collaboration}}{2016}]{Planck2016}
{Planck Collaboration} 2016, \mn@doi [\aap] {10.1051/0004-6361/201527101},
  \href {http://ukads.nottingham.ac.uk/abs/2016A%26A...594A...1P} {594, A1}

\bibitem[\protect\citeauthoryear{{Pontzen} \& {Governato}}{{Pontzen} \&
  {Governato}}{2012}]{Pontzen2012}
{Pontzen} A.,  {Governato} F.,  2012, \mn@doi [\mnras]
  {10.1111/j.1365-2966.2012.20571.x}, \href
  {https://ui.adsabs.harvard.edu/\#abs/2012MNRAS.421.3464P} {421, 3464}

\bibitem[\protect\citeauthoryear{{Puchwein}, {Baldi}  \& {Springel}}{{Puchwein}
  et~al.}{2013}]{Puchwein2013}
{Puchwein} E.,  {Baldi} M.,   {Springel} V.,  2013, \mn@doi [\mnras]
  {10.1093/mnras/stt1575}, \href
  {http://ukads.nottingham.ac.uk/abs/2013MNRAS.436..348P} {436, 348}

\bibitem[\protect\citeauthoryear{{Sakstein}}{{Sakstein}}{2013}]{Sakstein2013}
{Sakstein} J.,  2013, \mn@doi [\prd] {10.1103/PhysRevD.88.124013}, \href
  {http://ukads.nottingham.ac.uk/abs/2013PhRvD..88l4013S} {88, 124013}

\bibitem[\protect\citeauthoryear{{Sakstein} \& {Jain}}{{Sakstein} \&
  {Jain}}{2017}]{Sakstein2017}
{Sakstein} J.,  {Jain} B.,  2017, \mn@doi [Physical Review Letters]
  {10.1103/PhysRevLett.119.251303}, \href
  {http://adsabs.harvard.edu/abs/2017PhRvL.119y1303S} {119, 251303}

\bibitem[\protect\citeauthoryear{{Sawicki} \& {Bellini}}{{Sawicki} \&
  {Bellini}}{2015}]{Sawicki2015}
{Sawicki} I.,  {Bellini} E.,  2015, \mn@doi [\prd]
  {10.1103/PhysRevD.92.084061}, \href
  {http://ukads.nottingham.ac.uk/abs/2015PhRvD..92h4061S} {92, 084061}

\bibitem[\protect\citeauthoryear{{Schombert} \& {McGaugh}}{{Schombert} \&
  {McGaugh}}{2014}]{Schombert2014}
{Schombert} J.,  {McGaugh} S.,  2014, \mn@doi [\pasa] {10.1017/pasa.2014.32},
  \href {http://ukads.nottingham.ac.uk/abs/2014PASA...31...36S} {31, e036}

\bibitem[\protect\citeauthoryear{{Schwarz}}{{Schwarz}}{1978}]{Schwarz1978}
{Schwarz} G.,  1978, Annals of Statistics, \href
  {http://ukads.nottingham.ac.uk/abs/1978AnSta...6..461S} {6, 461}

\bibitem[\protect\citeauthoryear{{Vikram}, {Cabr{\'e}}, {Jain}  \& {Vand
  erPlas}}{{Vikram} et~al.}{2013}]{Vikram2013}
{Vikram} V.,  {Cabr{\'e}} A.,  {Jain} B.,   {Vand erPlas} J.~T.,  2013, \mn@doi
  [Journal of Cosmology and Astro-Particle Physics]
  {10.1088/1475-7516/2013/08/020}, \href
  {https://ui.adsabs.harvard.edu/abs/2013JCAP...08..020V} {2013, 020}

\bibitem[\protect\citeauthoryear{{Vikram}, {Sakstein}, {Davis}  \&
  {Neil}}{{Vikram} et~al.}{2018}]{Vikram2018}
{Vikram} V.,  {Sakstein} J.,  {Davis} C.,   {Neil} A.,  2018, \mn@doi [\prd]
  {10.1103/PhysRevD.97.104055}, \href
  {http://ukads.nottingham.ac.uk/abs/2018PhRvD..97j4055V} {97, 104055}

\bibitem[\protect\citeauthoryear{{Vogelsberger}, {Zavala}  \&
  {Loeb}}{{Vogelsberger} et~al.}{2012}]{Vogelsberger2012}
{Vogelsberger} M.,  {Zavala} J.,   {Loeb} A.,  2012, \mn@doi [\mnras]
  {10.1111/j.1365-2966.2012.21182.x}, \href
  {http://ukads.nottingham.ac.uk/abs/2012MNRAS.423.3740V} {423, 3740}

\bibitem[\protect\citeauthoryear{{Walker} \& {Pe{\~n}arrubia}}{{Walker} \&
  {Pe{\~n}arrubia}}{2011}]{Walker2011}
{Walker} M.~G.,  {Pe{\~n}arrubia} J.,  2011, \mn@doi [\apj]
  {10.1088/0004-637X/742/1/20}, \href
  {https://ui.adsabs.harvard.edu/\#abs/2011ApJ...742...20W} {742, 20}

\bibitem[\protect\citeauthoryear{{Xu}}{{Xu}}{2015}]{Xu2015}
{Xu} L.,  2015, \mn@doi [\prd] {10.1103/PhysRevD.91.063008}, \href
  {http://adsabs.harvard.edu/abs/2015PhRvD..91f3008X} {91, 063008}

\bibitem[\protect\citeauthoryear{{Zhao}, {Li}  \& {Koyama}}{{Zhao}
  et~al.}{2011a}]{Zhao2011a}
{Zhao} G.-B.,  {Li} B.,   {Koyama} K.,  2011a, \mn@doi [\prd]
  {10.1103/PhysRevD.83.044007}, \href
  {https://ui.adsabs.harvard.edu/abs/2011PhRvD..83d4007Z} {83, 044007}

\bibitem[\protect\citeauthoryear{{Zhao}, {Li}  \& {Koyama}}{{Zhao}
  et~al.}{2011b}]{Zhao2011b}
{Zhao} G.-B.,  {Li} B.,   {Koyama} K.,  2011b, \mn@doi [\prl]
  {10.1103/PhysRevLett.107.071303}, \href
  {https://ui.adsabs.harvard.edu/abs/2011PhRvL.107g1303Z} {107, 071303}

\bibitem[\protect\citeauthoryear{{de Vaucouleurs}, {de Vaucouleurs}, {Corwin},
  {Buta}, {Paturel}  \& {Fouque}}{{de Vaucouleurs}
  et~al.}{1991}]{deVaucouleurs1991}
{de Vaucouleurs} G.,  {de Vaucouleurs} A.,  {Corwin} Herold~G. J.,  {Buta}
  R.~J.,  {Paturel} G.,   {Fouque} P.,  1991, {Third Reference Catalogue of
  Bright Galaxies}

\makeatother
\end{thebibliography}

% APPENDIX
\appendix

\section{1D Approximation}
\label{A:1D2D_comparison}

\begin{figure*}
    \includegraphics{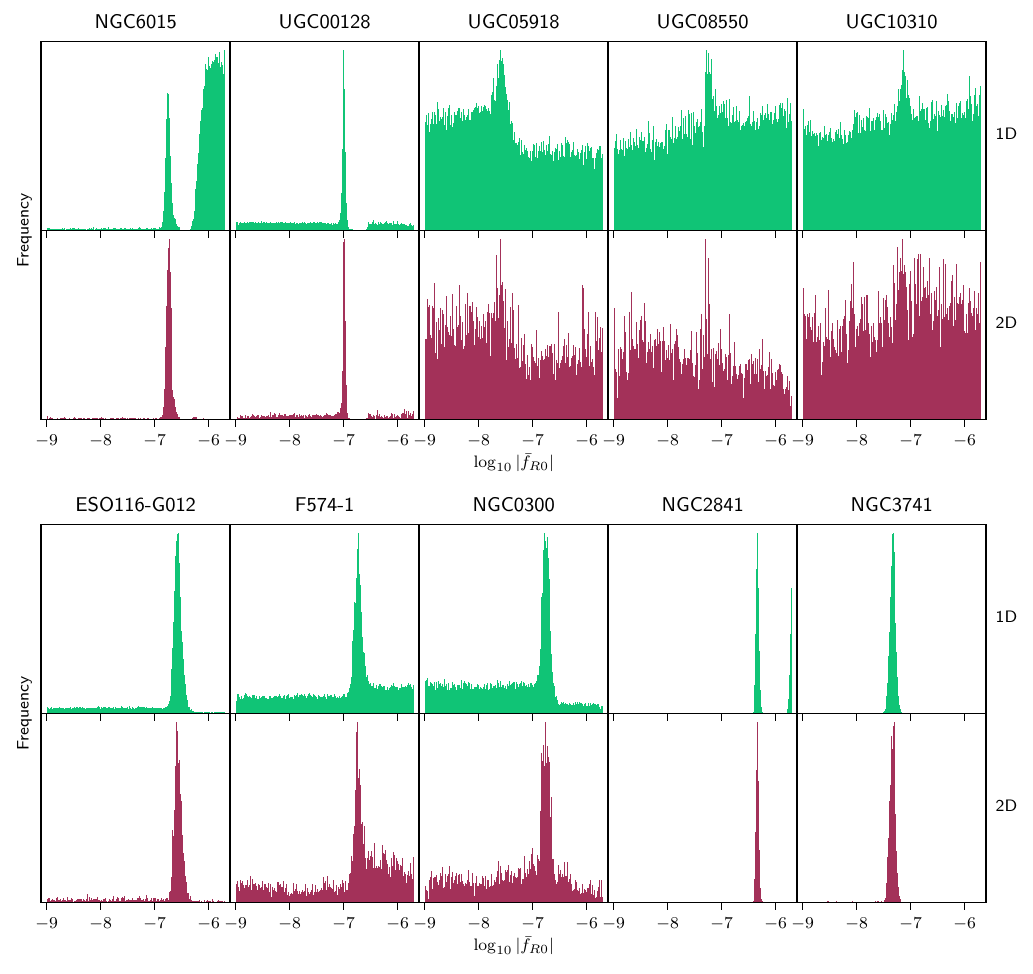}
    \centering
    \caption{Marginal posterior distributions of $\fR$ for 10 randomly selected galaxies, calculated with both the 1D solver used throughout this work (\textit{upper panels}), and the 2D solver described in this Appendix (\textit{lower panels}). The results of the two solvers are in remarkably good agreement, with the possible exception of the missing modes associated with NGC6015 and NGC2841, which are more likely to be due to convergence issues than the 1D approximation (each walker performed 5000 iterations in the 1D case, and 500 in the 2D case; see discussion in the text).}
    \label{F:2DTest}
\end{figure*}

As discussed in \S\ref{S:Methods:fRSolver}, when solving Equation (\ref{E:FieldEOM}) to calculate the scalar field, we assume spherical symmetry to reduce the computational cost. In Paper I, it was found that the scalar field typically takes a discoid shape in galaxies, so assuming spherical symmetry could potentially lead to our calculations differing from the `true' scalar field profiles, introducing error into our inferences for $\fR$. In particular, when a galaxy is partially screened, a given $\fR$ might correspond to an appreciably different location for the screening radius when the assumption of spherical symmetry is introduced.

To test the robustness of this 1D approximation, we compare some of our results with those that would have been obtained if the 1D solver was replaced with a 2D solver, in which the assumption of spherical symmetry has been replaced with the assumption of azimuthal symmetry. This 2D solver has been found to give galactic scalar field profiles in excellent agreement with those calculated by the full 3D solver in MG-GADGET.

In the 2D solver that we use for comparison, Equation~(\ref{E:EOM}) is discretised on a 2D grid in the spherical polar coordinates $r$ and $\theta$. As in the 1D case, we use logarithmically spaced grid cells for the coordinate $r$, i.e. the radial gridlines are equally spaced in $x \equiv \ln r$, with constant grid spacing $h_x$.

For the polar angle $\theta$, we adopt a linear grid, ranging from $\theta=0$ to $\theta=\pi$, with $n_\theta$ grid cells and constant grid spacing $h_\theta$. If $n_\theta$ is required to be an odd number, then the $\left(\frac{n_\theta+1}{2}\right)$\textsuperscript{th} grid cell corresponds to the disc plane. Through experimentation, it was found that $n_\theta=101$ gave sufficiently accurate results.

Defining
\begin{equation}\label{E:2DLij}
\mathcal{L}_{ij} \equiv \left(\nabla^2e^u\right)_{ij} +  \frac{1}{3c^2\bar{f}_R(a)}\left[\bar{R}\left(a\right)\left(1-e^{-\frac{u_{ij}}{2}}\right) + 8\pi G \delta \rho_{ij} \right],
\end{equation}
where $i$ and $j$ denote grid cells along the radial and polar directions respectively, we can rewrite Equation~(\ref{E:EOM}) as
\begin{equation}\label{E:2DDiscreteEOM}
\mathcal{L}_{ij}=0.
\end{equation}
This is then solved, with the same Newton-Gauss-Seidel technique, Equation (\ref{E:NGSUpdate}). In order to do this, we need discretised expressions for the Laplace operator on our grid, as well as the quantity $\frac{\partial\mathcal{L}_{ij}^n}{\partial u_{ij}^n}$. Writing $\sin\theta$ as $s$, the Laplace operator in our coordinates $x \equiv \ln r$ and $\theta$ is given by
\begin{equation}\label{E:2DLaplace}
\nabla^2f = \frac{1}{r^3}\frac{\partial}{\partial x}\left(r\frac{\partial f}{\partial x}\right) + \frac{1}{r^2 s}\frac{\partial}{\partial \theta}\left(s\frac{\partial f}{\partial \theta}\right),
\end{equation}
which is discretised as
\begin{multline}
\left(\nabla^2f\right)_{ij} = \frac{1}{r_i^3 h_x^2} \left( r_{i+\frac{1}{2}}(f_{i+1,j} - f_{ij}) - r_{i-\frac{1}{2}}(f_{ij} -f_{i-1,j})\right) \\ +\frac{1}{r_i^2 s_j h_\theta^2}\left(s_{j+\frac{1}{2}}(f_{i,j+1} - f_{ij}) - s_{j-\frac{1}{2}}(f_{ij} -f_{i,j-1})\right).
\end{multline}
Finally, the quantity $\frac{\partial\mathcal{L}_{ij}}{\partial u_{ij}}$ is given by
\begin{multline}
\frac{\partial\mathcal{L}_{ij}}{\partial u_{ij}} = \frac{\bar{R}}{6c^2\bar{f}_R}e^{-\frac{u_{ij}}{2}} \\ - e^{u_{ij}}\left[\frac{r_{i+\frac{1}{2}} + r_{i-\frac{1}{2}}}{r_i^3 h_x^2} + \frac{s_{j+\frac{1}{2}} + s_{j-\frac{1}{2}}}{r_i^2s_j h_\theta^2}\right].
\end{multline}

With a solution for $f_R$ everywhere, the fifth force is then given by Equation (\ref{E:a5}), evaluating the gradients of the scalar field in the disc plane, $\theta=\pi/2$. 

We take 10 randomly chosen galaxies from our sample, and rerun our MCMC analysis with this 2D solver for Model B, i.e. the $f(R)$ model with freely varying $\fR$, a single mass-to-light ratio, and NFW haloes. As in the 1D case, we use 30 walkers operating at 4 temperatures. With the 1D solver, the walkers each performed 5000 iterations after a burn-in period of 5000 iterations, but here we limit the computation to 500 iterations after a burn-in of 500, for reasons of computational cost. As a consequence, the convergence is more uncertain: whereas the Gelman-Rubin statistic $\mathcal{R}$ was previously confined to the region $|\mathcal{R}-1|<0.01$ for the vast majority of fits, the statistics for several of the galaxies are now in the range $0.01<|\mathcal{R}-1|<0.05$. 

Figure \ref{F:2DTest} shows the 10 marginal posterior distributions for $\fR$, calculated with 2D and 1D solvers. There is remarkably good agreement between the two solvers, demonstrating the validity of our 1D approximation. In two cases, NGC6015 and NGC2841, the posterior is bimodal for the 1D solver, but unimodal for 2D. However, as the missing mode in both cases falls within the fully unscreened regime, this is unlikely to be related to any difference between the scalar field solutions of the two solvers, but to the convergence issues described above, which have likely prevented the 2D MCMC from finding the second mode. 
%The walkers in the 2D MCMC have not been allowed a sufficient %number of iterations to find the second mode. 

% Don't change these lines
\bsp	% typesetting comment
\label{lastpage}
\end{document}